\documentclass[11pt]{article}
\usepackage{graphicx}
\usepackage{amssymb}
\usepackage{amscd}
\usepackage{mathrsfs}
\usepackage{longtable,lscape}
\usepackage{amsthm}
\usepackage{amsfonts}
\usepackage{amsmath}
\usepackage{bbm}
\usepackage{float}
\usepackage{url}
\usepackage{hyperref}
\usepackage[margin=0.5cm,font=footnotesize]{caption}
\usepackage{lineno}
\usepackage[round]{natbib}
\usepackage{appendix}
\usepackage{subfig}

\newcommand{\compl}{{\mathbb C}}
 
\begin{document}
\title{Human Perception as a Phenomenon of Quantization}
\author{Diederik Aerts and Jonito Aerts Argu\"elles  \vspace{0.5 cm} \\ 
        \normalsize\itshape
        Center Leo Apostel for Interdisciplinary Studies
        \\ 
        \normalsize\itshape
         Brussels Free University, Krijgskundestraat 33 \\ 
        \normalsize\itshape
         1160 Brussels, Belgium \\
        \normalsize
        E-Mails: \url{diraerts@vub.ac.be, diraerts@gmail.com},
        \\ \url{jonitoarguelles@gmail.com},
              	\\
              }
\date{}
\maketitle
\begin{abstract}
\noindent
For two decades, the formalism of quantum mechanics has been successfully used to describe human decision processes, situations of heuristic reasoning, and the contextuality of concepts and their combinations. The phenomenon of `categorical perception' has put us on  track to find a possible deeper cause of the presence of this quantum structure in human cognition. Thus, we show that in an archetype of human perception consisting of the reconciliation of a bottom up stimulus with a top down cognitive expectation pattern, there arises the typical warping of categorical perception, where groups of stimuli clump together to form quanta, which move away from each other and lead to a discretization of a dimension. The individual concepts, which are these quanta, can be modeled by a quantum prototype theory with the square of the absolute value of a corresponding Schr\"odinger wave function as the fuzzy prototype structure, and the superposition of two such wave functions accounts for the interference pattern that occurs when these concepts are combined. Using a simple quantum measurement model, we analyze this archetype of human perception, provide an overview of the experimental evidence base for categorical perception with the phenomenon of warping leading to quantization, and illustrate our analyses with two examples worked out in detail.
\end{abstract}

\section{Introduction \label{introduction}}
The identification of the presence of quantum structures in human cognition, and their importance for the development of performant quantum structure-based models for bounded rationality situations and human decisions, led to a field of research now called `quantum cognition' \citep{aertsaerts1995,gaboraaerts2002,bruzacole2005,aertsgabora2005a,aertsgabora2005b,busemeyeretal2006,aerts2009a,aerts2009b,bruzagabora2009,aertssozzo2011,aertsetal2012,busemeyerbruza2012,havenkhrennikov2013,dallachiaraetal2015,pothosetal2015,blutnerbeimgraben2016,moreirawichert2016,gaborakitto2017,surov2021,pothosbusemeyer2022}.
Parallel to the emergence of the research field of quantum cognition, quantum structures were identified in computer science. Their use  in information retrieval and natural language processing especially lead to important results, and also brought a new research domain to life called `quantum information science' \citep{aertsczachor2004,vanrijsbergen2004,widdows2004,piwowarskietal2010,songetal2010,melucci2015,aertsetal2019}.

One of the authors, using an approach that belongs to this field of research `quantum information science', studied the presence and importance of two quantum structures, namely, `superposition' and `entanglement', in human visual perception, by using `Google Images' as a dataset \citep{aertsarguelles2018}. More specifically, the following was demonstrated. When collecting the  images that appear with the search term `child', and also collecting the images that appear with the search term `mother', and then collecting the images that appear with the two search terms `child' and `mother', an emergence takes place. Indeed, many images appear of a mother holding a child in her arms. These specific images are much less present in the first two collections. A similar emergence occurs with another example, namely when images are collected that turn up with the search term `glass', and then images that turn up with the search term `bottle', and then images that turn up with the two search terms `glass' and `bottle'. The emergence that takes place here are images in which the glass contains drink from the bottle and in which possibly drinking itself also takes place. It is the quantum structure of `superposition' that is at work here, bringing about this emergence as an interference effect. We know this because the situation is very similar to the `pet-fish problem', or the `guppy effect', in concept theory \citep{oshersonsmith1981}. That guppy effect is well expressed in the statement, ``A guppy is neither a typical pet nor a typical fish, but it is a typical pet-fish''. The similarity can be easily understood as follows, ``A~mother with child on her arm is neither a typical image of a mother nor a typical image of a child, but it is a typical image of a mother and a child''. Equally, ``A glass with drink in it from a bottle is not a typical image of a glass, nor is it a typical image of a bottle, but it is a typical image of a glass and a bottle''. That the guppy effect can be modeled using the quantum structure we call superposition, and the data then appear in the model as a result of interference, has been demonstrated by one of the authors and  his collaborators and analyzed in various ways in past publications \citep{gaboraaerts2002,aertsgabora2005a,aertsgabora2005b,aertsetal2012}. The same method can be used to model these experimental results about mother and child, and glass and bottle, in \citep{aertsarguelles2018}. 

These are the results regarding `superposition', but  `entanglement' was also identified in visual perception using Google Images as a data set in \citep{aertsarguelles2018}. For this, the sentence, `The animal acts', was used, which was originally considered to demonstrate `entanglement' in quantum cognition using a psychological experiment \citep{aertssozzo2011}. More specifically, the Clauser Horne Shimony Holt inequalities \citep{clauseretal1969}, a version of Bell's inequalities \citep{bell1964}, are violated by fitting the fractions in specific collections of Google Images that turn up for specific search terms into these inequalities in the prescribed way \citep{aertsarguelles2018}.

These results, obtained by data originating in human visual perception, namely, Google Images, demonstrate the presence of quantum structures in human visual perception. However, the connections within the considered elements of human visual perception, namely, Google Images, are likely to be located more in the cognitive and meaning-bearing phase of visual perception than in the first-line perceptual phase.
Therefore, in this article, we would like to pay attention to visual perception that takes place in this more primitive first-line phase and quantum structures that would be present there. A specific situation in visual perception, namely, the bi-stability that occurs when viewing figures drawn on a two-dimensional background that we nevertheless visually reconstruct into `seeing three dimensional entities', of which the `Necker cube' is the archetypal example, was studied within the quantum cognition approach. The presence of quantum structure was investigated and convincingly demonstrated \citep{conteetal2009,atmanspacherfilk2010}.

In this article, however, our focus is not on demonstrating the presence of quantum structures in human perception in specific cases, such as the one mentioned above, for the visual perception of Necker-cube-like figures. Rather, we wish to show that human perception in its very general form already carries the same structure and dynamics as that of the Necker cube-like figures, and that as a consequence, with great plausibility, the presence of an underlying quantum structure can be hypothesized. We also identify how, additionally to the specific quantum-like structural and dynamic nature of human perception, there occurs a phenomenon called `categorical perception', which from the quantum structure underlying the basic level of human perception, gives rise to outright quantization, more precisely, to the discretization of the categorical precepts. Furthermore, we show how the quanta formed by the mechanism of categorical perception can be well described on an individual basis by a prototype theoretical model, namely, by a fuzzy structure of a prototype surrounded by a cloud of its exemplars.
This is very analogous to how physicists describe the quanta of a physical phenomenon, e.g., a beam of light, by wave packets, except that, applying quantum mechanics, physicists mathematically represent the wave packets by complex functions, where only the squares of their absolute values fill in the fuzzy structure. We analyzed in earlier work how this use of the mathematically deeper layer of complex functions for the prototype structure can also be introduced for concepts, and then allows one to describe also the combination of concepts, for which generally no prototype exists, by the quantum superposition, namely, the renormalized sum, of these complex functions. Interference is then the phenomenon that occurs at the level of the fuzzy structure, and allows one to model the combination of two concepts. Thus, our earlier work joins the present one, and in this way we get a model for a complete evolution of the formation of human concepts and human language. Starting from general human perception, moving on to the warping with which categorical perception introduces quantization and the prototype model for the quanta, and ending up at the quantum model for the combination of concepts.
We already mentioned how the research in \citep{aertsarguelles2018} was inspired by earlier research about the guppy effect in concept theory in which one of the authors was directly involved \citep{gaboraaerts2002,aertsgabora2005a,aertsgabora2005b,aertsetal2012}. In the present article, we will also partly rely on insights obtained in this earlier research. In this respect, we mention that `the guppy effect' in concept theory is very similar to well-specified effects that appear in human decisions. We think here of the phenomena that were widely studied by Amor Tversky and collaborators in the 1980s and the 1990s, such as the conjunction fallacy and the disjunction effect \citep{shafir2018}. It has been one of the important undertakings in quantum cognition to show that the effects studied by Tversky and collaborators can be modeled by quantum decision approaches with better results than by classical decision approaches \citep{aerts2009a,busemeyerbruza2012,havenkhrennikov2013,pothosbusemeyer2022}. 

In Section \ref{measurementfluctuations}, we introduce a quantum measurement model, using only elements that allow one to easily imagine what is happening in a quantum measurement. We show that quantum probabilities can be derived from the simple geometry of the model itself. We wish to use the model for the simple insights it provides into aspects of the quantum measurement process, for example, the presence of fluctuations on the interactions between measuring apparatus and the entity being measured, which are not easy to grasp.

In Section \ref{quantumperception}, we introduce the quantum mechanical description in a two-dimensional complex Hilbert space for the measurement model. We show that it is a model for a qubit in the Bloch sphere with appended measurement facility. We analyze the contextuality and the purity and mixedness of states and their von Neumann entropy. We also introduce the basic experiment on the nature of human perception and make a preliminary connection to quantum superposition.

In Section \ref{categoricalquantum}, we introduce the mechanism of categorical perception, and describe the way it produces a warping of the space of perceptions. This warping leads to a form of quantization to finally culminate in the structural proposal of prototype theory. We analyze how this prototype theory runs into problems when composite concepts are examined, and how the underlying quantum structure puts forward a solution to this in the form of interference. The quantum prototype theory we develop is dynamically contextual and rooted in operational quantum axiomatics, and is therefore able to model the combination of concepts with interference as a manifestation of this underlying quantum structure. We illustrate this with two examples worked out in detail, the concepts {\it {Fruits}} and {\it Vegetables} being combined into the concept {\it Fruits or Vegetables}, and the concepts {\it Furniture} and {\it Household Appliances} being combined into the concept {\it Furniture and Household appliances} (we will denote `concepts' in italics and with an initial capital letter, the notation we applied in our earlier works). 

\section{Fluctuations in Measurement Interactions \label{measurementfluctuations}}

In non-relativistic quantum mechanics, there are two types of change of state. There is the continuous change in time that describes the evolution of a quantum entity when not interacting directly with it, which is well modeled by the Schr\"odinger equation. Additionally, there is the discrete change of state, which represents what happens when interacting directly with a quantum entity, such as when a measurement is being performed. The first type of change is deterministic, whereas the second type of change is probabilistic. The existence of these two types of change of state has been considered problematic by many physicists and philosophers, and through time, not unrelated to the context in which it was being reflected upon, has been referred to as `the measurement problem in quantum mechanics'. Many different interpretations of quantum mechanics originate at this intersection of what is called `the measurement problem' of quantum mechanics, representing different views on this problem \citep{howard2004,camilleri2009,kent2010,bassietal2013, goldstein2022}.

In the years before the period arrived where the testing of Bell's inequalities prepared the physics community for accepting the reality of the quantum phenomenon of entanglement, one of the authors was working on an axiomatic approach to quantum mechanics~\citep{aerts1982,aerts1992}, and from that vantage point also elaborated a model for a measurement in quantum mechanics \citep{aerts1986}. In the model, the quantum probabilities are caused by the presence of fluctuations in the interactions of the measuring apparatus with the quantum entity on which is measured. More specifically, this means that repeating the same measurement is more complicated than is generally thought, namely, every time a measurement is supposedly repeated, there is nevertheless a difference in the interactions that take place between the measuring apparatus and the entity on which is measured. In this model, it is the presence of these differences between supposedly identical measurements and the statistics with which these differences occur in the realm of the interactions between the measuring apparatus and the quantum entity on which is measured that is the origin of the quantum probabilities.

\begin{figure}[h!]
\begin{center}
\includegraphics[width=7cm]{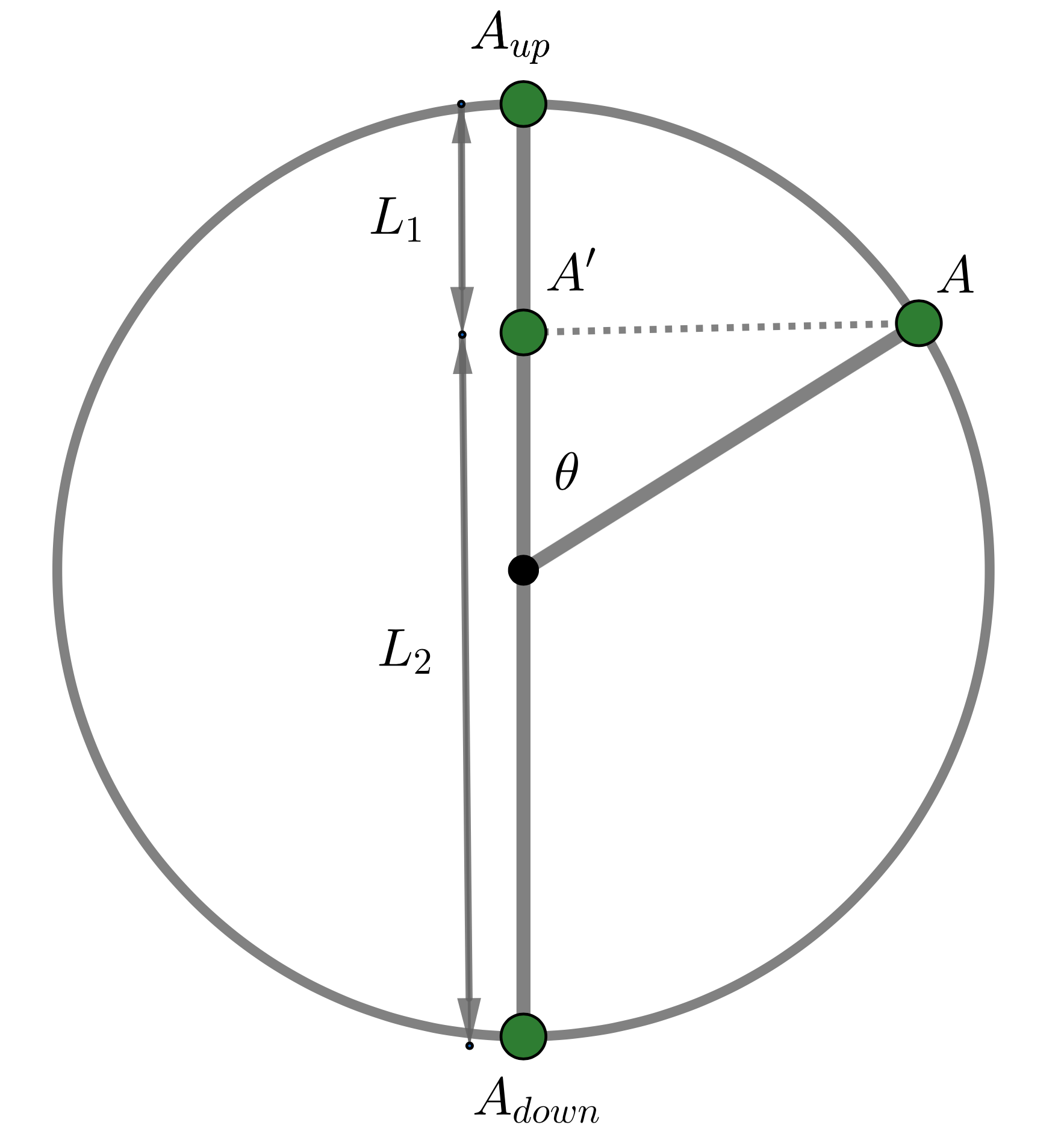}
\end{center}
\caption{An example of a situation with fluctuations in the interaction between a measurement apparatus and the entity to be measured, which we call the `elastic sphere model'. A little ball is placed in a point $A$ on a sphere, and this represents the state of the entity before the measurement. On a central axis of the sphere, between the points $A_{down}$ and $A_{up}$, on the surface of the sphere, is placed an elastic and a first phase of the measurement consisting of the ball falling orthogonally to the elastic and sticking to it at point $A'$. The second phase of the measurement consists of the elastic breaking at one of its points randomly and uniformly, hence such that the probability of breaking in a piece is proportional to the length of this piece. As a consequence, the ball is moved upwards, ending up at point $A_{up}$, or downwards, ending up at point $A_{down}$, depending on whether the elastic breaks in the part below $A'$ or the part above $A'$.}
\label{ElasticSphereModelfigure}
\end{figure}

What we are writing here seems complicated at first sight, and since it is important to understand it well given what we wish to put forward in this article, we include a simple example where each aspect is illustrated. The example is shown in Figure \ref{ElasticSphereModelfigure} and consists of a piece of elastic stretched between points $A_{down}$ and $A_{up}$ on the centerline of a sphere. On the surface of the sphere is a small ball that sticks to the sphere at a point $A$. This ball in a point represents the state of the quantum entity in our example. The measurement occurs as follows. First, the ball falls orthogonally on the elastic and sticks to it at point $A'$ where this orthogonal fall brings it. Then, the elastic breaks randomly at one of its points. If the breaking point is below point $A'$, then the unbroken part of the elastic pulls the ball up so that it ends up at point $A_{up}$. However, if the breaking point is above point $A'$, then the ball is pulled down by the unbroken part of the elastic, and ends up at point $A_{down}$. Thus, the whole of the measurement results in the ball from point $A$ ending up at one of two points, $A_{up}$ or $A_{down}$.

\begin{figure}[h!]
\begin{center}
\includegraphics[width=10cm]{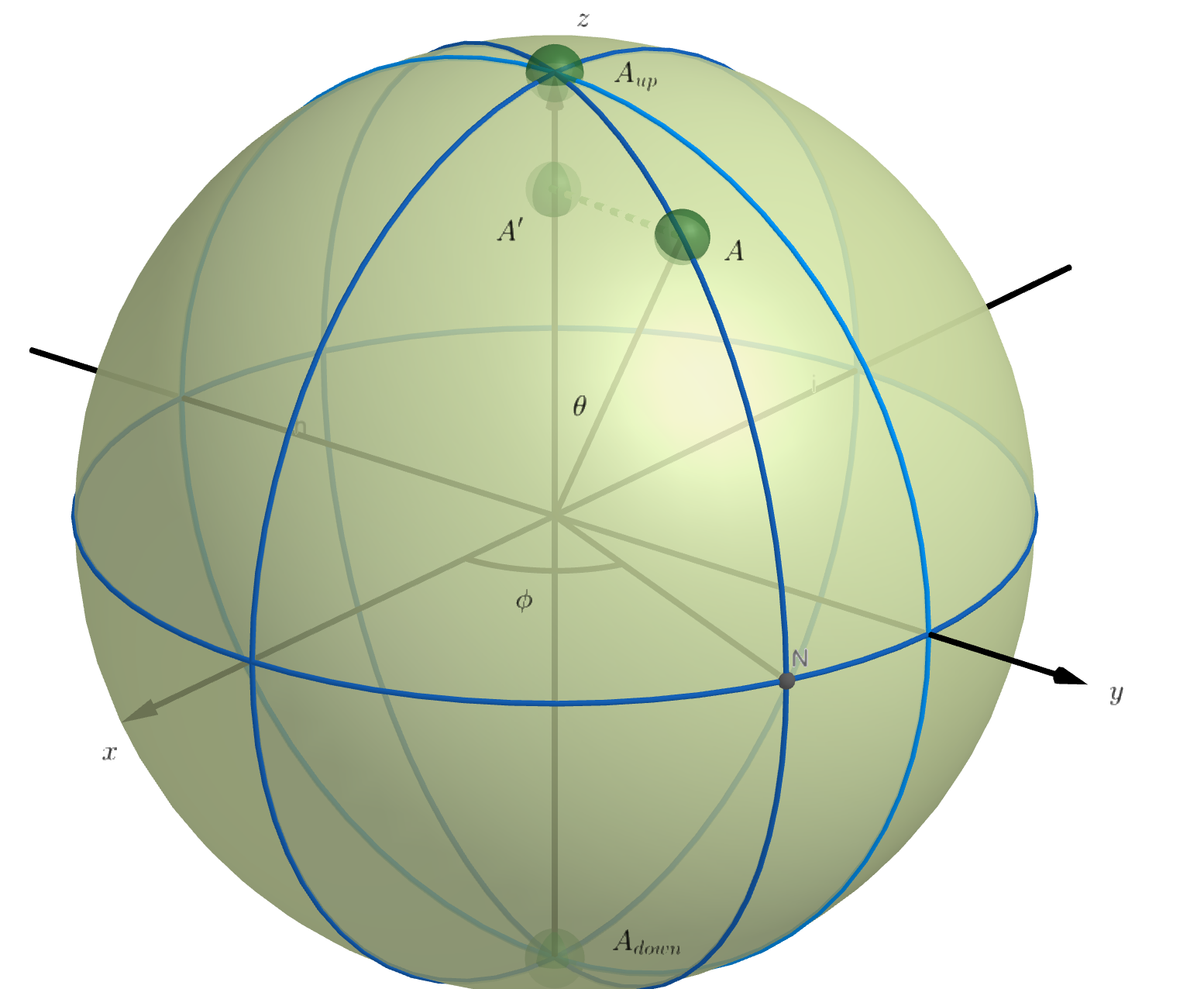}
\end{center} 
\caption{{A three} dimensional representation of the elastic sphere model, revealing it to be a quantum Bloch sphere model extended with an explicitation of the measurement procedure. The little ball is at point $A$ with spherical coordinates $(\rho, \theta, \phi)$ and Hilbert space density operator coordinates $D_{(\rho, \theta, \phi)}$. The process of measurement proceeds as described in detail in Figure \ref{ElasticSphereModelfigure} by the ball falling orthogonally to the elastic and sticking to it at point $A'$. Then, the elastic breaks uniformly in one of its points and pulls the ball upwards to end up in $A_{up}$ or downwards to end up in $A_{down}$.}
\label{ElasticSphereModel3Dfigure}
\end{figure}

Now suppose that the elastic possesses a uniform break pattern, which means that the point at which the elastic breaks randomly will lie within a certain interval of the elastic with a probability proportional to the length of that interval. We can then calculate the probabilities $P(A \mapsto A_{up})$ and $P(A \mapsto A_{down})$ with which the measurement will carry the ball from $A$ to $A_{up}$ or to $A_{down}$, and in this way will change probabilistically the state of the quantum entity, from the simple geometry of the configuration. Note indeed that if $r$ is the radius of the sphere, we have that the length $L_2$ of the elastic under point $A'$ equals $r+r\cos\theta$ and the length $L_1$ of the elastic above point $A'$ equals $r-r\cos\theta$, which hence, taking into account the uniform nature of the breaking pattern of the elastic, gives us
\begin{eqnarray} \label{qubit01}
P(A \mapsto A_{up}) &=& {r+r\cos\theta \over 2r} = {1 \over 2}(1+\cos\theta) = \cos^2 {\theta \over 2} \\ \label{qubit02}
P(A \mapsto A_{down}) &=& {r-r\cos\theta \over 2r} = {1 \over 2}(1-\cos\theta) = \sin^2 {\theta \over 2}
\end{eqnarray}
Some have probably understood by now that this elastic sphere model is a model for a quantum qubit, the probabilities (\ref{qubit01}) and (\ref{qubit02}) indeed match the quantum probabilities of the spin of a spin 1/2 quantum particle. The ball at point $A$ is then in a state for the spin of an angle $\theta$ with the $z$-axis, and the elastic lies on this $z$-axis.
In Figure \ref{ElasticSphereModel3Dfigure}, we provide a three-dimensional representation of the situation. 
Some have probably also seen that the sphere and the points on its surface where $A$ can be located is none other than the so-called Bloch sphere, a well-known model for the spin of a spin 1/2 quantum particle or a qubit. However, we have extended this Bloch sphere to include a model for the measurement itself, and that is where the elastic comes in, on the one hand, with the orthogonal falling of the ball from point $A$ to point $A'$, and on the other hand, with the uniform random breaking of the elastic at one of its points.

At the time when one of the authors developed this model \citep{aerts1986}, the focus in connection with quantum mechanics was different than it is today. One asked fundamental questions of a more structural nature, in connection with, e.g., `the structure of the set of properties of a quantum entity with respect to a classical entity'---this structure being non-Boolean for the quantum entity while Boolean for the classical entity---or `the structure of the quantum probability model with respect to a classical probability model'---this structure being non-Kolmogorovian for a quantum entity and Kolmogorovian for a classical entity. As such, this model was conceived in the 1980s to answer questions posed at that time. Let us briefly elaborate on one of these questions, because the answer we can formulate to it is interesting and important for the problem we are dealing with here, even for the research domain of quantum cognition in itself, and more specifically, for its scientific status. In connection with probabilities, for example, one may ask whether they can be due to a lack of knowledge about an underlying structure that may even be deterministic, since, if this were the case for the probabilities occurring in quantum mechanics, one could then introduce additional variables, which would eliminate this lack of knowledge, by better and even fully describing that underlying reality. This structural question was called `the hidden variable problem' of quantum mechanics.
Several proofs existed in the 1980s of the impossibility of obtaining the probabilities of quantum mechanics as a consequence of the presence of hidden variables, among which von Neumann's proof was probably the best known \citep{vonneumann1932}. In other words, these no-go theorems for hidden variables showed that the quantum probabilities were not due to a lack of knowledge about an underlying and not (yet) known reality.
In the qubit model of our example in Figures \ref{ElasticSphereModelfigure} and \ref{ElasticSphereModel3Dfigure}, we can note that if we allow the point where the elastic breaks to be introduced as an additional variable, what takes place during the measurement process no longer contains probabilities, and thus is deterministic, when this variable is known. But, we have clearly stated that our entity is determined by the location on the sphere of the ball, i.e., by the location of point $A$, and variables that describe where the elastic breaks are thus not variables that give us more knowledge about the state of the entity, and thus are not variables of the entity. That means our model escapes the no-go theorems about hidden variables, and demonstrating that was the main goal in the 1980s of the measurement model we present here. Indeed, each of the points where the elastic can break corresponds to a well-defined interaction between the entity at point $A$ and the measuring apparatus, which is the whole of the sphere and the elastic. It is these various possible interactions between the entity and the measuring apparatus that we have called, in the more abstract formulation above, `fluctuations in the interactions between measuring apparatus and entity'. For the present article we are less interested in the significance of this model for structural questions concerning, e.g., the nature of the quantum probabilities, although, one result is worth mentioning. It can be proven (see page 205 of \citep{aerts1986}) that the set of transition probabilities contained in the measurement model, considered as conditional probabilities, cannot be described by a classical Kolmogorovian probability model \citep{gudderzanghi1984}.

What we are particularly interested in for our present article is the fine structure of the mechanism in the measurement in the example shown in Figures \ref{ElasticSphereModelfigure} and \ref{ElasticSphereModel3Dfigure}. And before we continue our analysis, we wish to note the following. The example depicts a quantum measurement with two final states $A_{up}$ and $A_{down}$, and thus a quantum measurement with two outcomes. However, in later years, in collaboration of one of the authors with Massimiliano Sassoli de Bianchi, it was shown that a similar model can be built for a quantum mechanical measurement with an arbitrary number $n$ of final states, and thus an arbitrary number $n$ of outcomes \citep{aertssassolidebianchi2014}.
We will not explicitly describe these higher dimensional extended Boch sphere quantum measurement models in the present article, but mention that the simple geometric properties of the configurations lead to exactly the quantum probabilities in a completely similar way than this is the case for this two-dimensional quantum measurement model. 
The details of the analysis that we will now present for the two-dimensional quantum measurement model can also be made for the higher dimensional quantum measurement models in a similar way.
Of course, that we find exact quantum probabilities in these measurement models also depends on a certain symmetry that we introduced at the measurement level, namely, that the elastic breaks with a randomness that is `uniformly' distributed. If the elastic does not possess this symmetry property, we will still find probabilities that do not fit a classical probability model, but they will not be perfect quantum probabilities. Hence, a complex Hilbert space model will not be able to represent them since they will be rather quantum-like instead of pure quantum. More general quantum formalisms than this of standard quantum mechanics in a complex Hilbert space, and not based on vectors representing states, can then be used \citep{aerts1982,aerts1992,gudderzanghi1984}. We return to this briefly in Section \ref{categoricalquantum} when we mention the SCoP model for concepts. It would lead us too far away from the subject of this article to elaborate on it now, but it is something we do plan for future work.

Now, before continuing our analysis of this measurement model, let us come to that general statement about the scientific status of quantum cognition, as we promised. It is frequently argued by other researchers in quantum cognition that we are not interested in finding the presence of `quantum in cognition' that would be due to the some microscopic activity of quantum physics in the human brain (e.g., sometimes microtubules are put forward as a possible
seat for such microscopic quantum physics activities by researchers who are in search of this \citep{hameroffpenrose2014}). 
We agree with this attitude amongst the researchers working in quantum cognition, and want to explicitly reflect on this using the quantum model for a qubit that we just introduced. Suppose for a moment that we hide the `elastic sphere model' in a black box, and then have a technician place the elastic and the ball in ever-changing directions and angles, and for each of these setups the data are collected, i.e., either an outcome in $A_{down}$ of that particular setup or an outcome in $A_{up}$ of that particular setup. The technician may see the interior of the black box, because he or she has to organize the different setups, but for the researcher who will analyze the data, the entire apparatus, which is the elastic sphere model that we just introduced, is hidden inside the black box. It is clear that the data, after a thorough analysis, will show that they are equivalent to data obtained after quantum measurements on a qubit in states corresponding to the different setups. There is no doubt that a material realization of the elastic sphere model, i.e., with true elastics, and a sphere made of a substance that allows the experiments to be truly performed, would be macroscopic. This means that the actual possibility of the existence of such a material elastic sphere model demonstrates that the quantum probability model of a qubit can realize itself in the macroscopic world. There is no need for the presence of a microscopic element in the human brain where the probability model of a qubit would realize itself from quantum physics applicable to this microscopic element of the human brain. The analysis developed in this section, in addition to being directed toward the researchers who are looking for such a microscopic element in the human brain, is even more importantly directed toward the skeptics of `quantum and human cognition'. Indeed, it is primarily these skeptics who believe that `quantum and human cognition' are only meaningful to even consider from a scientific perspective `if' there is such a microscopic element in the human brain that causes the quantum structure to emerge. The elastic sphere model of a qubit, which is also a model of the spin of a spin 1/2 quantum particle, and the fact that a material realization can be made of it, from which, while hidden in a black box, data can be collected, shows to these skeptics that there is no problem with macroscopically realizing the full probability model of a qubit, including the set of states and its quantum structure. Some skeptics will note that it is primarily the entanglement-style quantum mechanical assembly of several qubits that is important, and may be willing to be convinced that a single loose qubit is indeed realized with this elastic sphere model. This is then where the violation of Bell inequalities comes in, and indeed, this is as important an aspect of quantum mechanics as the dynamics and probability structure of a single qubit. That such a violation of Bell's inequalities occurs when one combines concepts \citep{aertssozzo2011} is then interesting for them to look at more closely.

\section{Quantum Measurement and Human Perception \label{quantumperception}}
What we wish to analyze now, in the example of Figures \ref{ElasticSphereModelfigure} and \ref{ElasticSphereModel3Dfigure}, is `how this example, and the way it describes a measurement in detail', could be used for measurements in~psychology.

First, we want to distinguish the two phases of the measurement more precisely than we already did. The first phase consists of point $A$ falling orthogonally to the elastic between the two points $A_{down}$ and $A_{up}$. The entity in state $A$ changes as a result and ends up in state $A'$.
We note that, observing the Bloch representation, the state of the quantum entity in point $A$ is a `pure state', whereas the state of the quantum entity in point $A'$ is a `mixed state'. A pure state means, among other things, that the von Neumann entropy is equal to zero. The mixed state in point $A'$ is not just any mixed state, it is a convex combination of the states in points $A_{down}$ and $A_{up}$. This means that the specific mixed state in point $A'$ is a `classical' mixture of the two possible final states, this one in $A_{up}$ and this one in $A_{down}$, which are both pure states. Before we forget to mention this, the pure state in $A$ is a quantum superposition of the two pure states in $A_{down}$ and $A_{up}$.
Let us make this explicit with the standard calculations of the spin of a spin 1/2 quantum particle, or a qubit.

The Hilbert space we consider is $\compl^2$.
The state of the spin of the spin 1/2 quantum particle in the direction
$(\theta, \phi)$, is represented by the point 
$(\sin\theta\sin\phi, \sin\theta\cos\phi, \cos\theta)$, and that is also where in Figure \ref{ElasticSphereModel3Dfigure} the ball is located in point $A$. In the two-dimensional complex Hilbert space, this point $A$ is now represented by the vector $|\theta,\phi\rangle = (\cos{\theta \over 2} e^{-i{\phi \over 2}}, \sin{\theta \over 2} e^{i{\phi \over 2}}) \in \compl^2$, the angles $\theta$ and $\phi$ are indicated in Figure \ref{ElasticSphereModel3Dfigure}, and are actually the angles of the classical spherical coordinates $(\rho, \theta, \phi)$ of point $A$, with hence coordinates in the three dimensional real space given by $(\rho\sin\theta\sin\phi,\rho\sin\theta\cos\phi,\rho\cos\theta)$.
The sphere in Figure \ref{ElasticSphereModel3Dfigure} has radius~1, and hence $\rho=1$ for $A$, which is situated on the surface of the sphere. Points of the interior of the Bloch sphere correspond to density operators, hence, let us calculate the density operator corresponding to point $A'$. A pure state corresponding to point $A_{up}$, with coordinates (0,0,1), is $|0, \phi\rangle = (e^{-i{\phi \over 2}}, 0)$, and a pure state corresponding to point $A_{down}$, with coordinates (0,0,-1), is $|\pi, \phi\rangle = (0, e^{i{\phi \over 2}})$. We will denote a density operator corresponding to the point with spherical coordinates $(\rho, \theta, \phi)$, $\rho \in [0,1]$, $\theta \in [0,\pi]$, $\phi \in [0, 2\pi]$ as $D_{(\rho, \theta, \phi)}$. Let us first calculate the density operators corresponding to the three pure states $A$, $A_{up}$ and $A_{down}$. We have
\begin{eqnarray}
D_{(1,\theta,\phi)} &=& |\theta, \phi\rangle \langle \theta, \phi|
= \begin{pmatrix}
\cos{\theta \over 2} e^{i{\phi \over 2}} \\
\sin{\theta \over 2} e^{-i{\phi \over 2}}
\end{pmatrix}
\begin{pmatrix}
\cos{\theta \over 2} e^{-i{\phi \over 2}} & \sin{\theta \over 2} e^{i{\phi \over 2}}
\end{pmatrix} \nonumber \\
&=& \begin{pmatrix}
\cos^2{\theta \over 2} & \cos{\theta \over 2}\sin{\theta \over 2}e^{i\phi} \\
\cos{\theta \over 2}\sin{\theta \over 2}e^{-i\phi} & \sin^2{\theta \over 2}
\end{pmatrix} \\ \label{theta0}
D_{(1,0,\phi)} &=& \begin{pmatrix}
\cos^2{\theta \over 2} & \cos{\theta \over 2}\sin{\theta \over 2}e^{i\phi} \\
\cos{\theta \over 2}\sin{\theta \over 2}e^{-i\phi} & \sin^2{\theta \over 2}
\end{pmatrix}_{\theta=0,\phi}
= \begin{pmatrix}
1 & 0 \\
0 & 0
\end{pmatrix}
\\ \label{thetapi}
D_{(1,\pi,\phi)} &=&\begin{pmatrix}
\cos^2{\theta \over 2} & \cos{\theta \over 2}\sin{\theta \over 2}e^{i\phi} \\
\cos{\theta \over 2}\sin{\theta \over 2}e^{-i\phi} & \sin^2{\theta \over 2}
\end{pmatrix}_{\theta=0,\phi}
= \begin{pmatrix}
0 & 0 \\
0 & 1
\end{pmatrix}
\end{eqnarray}
Let us now calculate the density operator $D_{A'}$ representing the state of the entity when it is in point $A'$, as in Figure \ref{ElasticSphereModel3Dfigure}. We remark that $A'$ lies on the line between $A_{down}$ and $A_{up}$ sticking to the elastic, which is stretched between $A_{down}$ and $A_{up}$ on this line. Making use of a general property of the set of all density operators, i.e., that it is a set closed by convex combination, we know that $D_{A'}$ is a convex combination of $D_{(1,\pi,\phi)}$ and $D_{(1,0,\phi)}$, which~gives

\begin{eqnarray} \label{convex01}
D_{A'} = \lambda \begin{pmatrix}
0 & 0 \\
0 & 1
\end{pmatrix}
+ (1-\lambda) \begin{pmatrix}
1 & 0 \\
0 & 0
\end{pmatrix}
= \begin{pmatrix}
1-\lambda & 0 \\
0 & \lambda
\end{pmatrix}
\end{eqnarray}
for $\lambda \in [0, 1]$. From Figure \ref{ElasticSphereModel3Dfigure}, given that $A'$ is obtained by projecting orthogonally to the line between $A_{down}$ and $A_{up}$, we have that $A'$ lies on the line between $A$ and the point with spherical coordinates $(1,\theta,\phi+\pi)$, to which corresponds the density operator
\begin{eqnarray}
D_{(1,\theta,\phi+\pi)} &=&
\begin{pmatrix}
\cos^2{\theta \over 2} & \cos{\theta \over 2}\sin{\theta \over 2}e^{i(\phi+\pi)} \\
\cos{\theta \over 2}\sin{\theta \over 2}e^{-i(\phi+\pi)} & \sin^2{\theta \over 2}
\end{pmatrix} \nonumber \\
&=&
\begin{pmatrix}
\cos^2{\theta \over 2} & -\cos{\theta \over 2}\sin{\theta \over 2}e^{i\phi} \\
-\cos{\theta \over 2}\sin{\theta \over 2}e^{-i\phi} & \sin^2{\theta \over 2}
\end{pmatrix}
\end{eqnarray}
This means that we have
\begin{eqnarray} \label{convex02}
D_{A'} = \mu D_{(1,\theta,\phi)} + (1-\mu) D_{(1,\theta,\phi+\pi)}
\end{eqnarray}
for $\mu \in [0, 1]$. From (\ref{convex01}) and (\ref{convex02}), it follows that we must have
\begin{eqnarray}
&&\mu \cos{\theta \over 2}\sin{\theta \over 2}e^{-i\phi} - (1-\mu) \cos{\theta \over 2}\sin{\theta \over 2}e^{-i\phi} =0 \nonumber \\
&&\Leftrightarrow
\mu - (1-\mu) = 0 \nonumber \\
&& \Leftrightarrow
\mu = {1 \over 2}
\end{eqnarray}
and
\begin{eqnarray}
\lambda = \sin^2{\theta \over 2}
\end{eqnarray}
This gives us
\begin{eqnarray}
D_{A'} &=& \begin{pmatrix}
\cos^2{\theta \over 2} & 0 \\
0 & \sin^2{\theta \over 2}
\end{pmatrix} =
\cos^2{\theta \over 2}
\begin{pmatrix}
1 & 0 \\
0 & 0
\end{pmatrix}
+ \sin^2{\theta \over 2}
\begin{pmatrix}
0 & 0 \\
0 & 1
\end{pmatrix} \nonumber \\
&=& \cos^2{\theta \over 2} D_{A_{up}} + \sin^2{\theta \over 2} D_{A_{down}}
\end{eqnarray}
Let us interpret the results of these calculations. First, we note again that the states in which the entity, the ball, is in, when being in points $A$, $A_{down}$, and $A_{up}$, are pure states. This means that the von Neumann entropies of these states are equal to zero. The state in which the entity, the ball, is in, being in point $A'$, is not a pure state. Let us calculate its von Neumann entropy $S_{A'}$. To do so, we use the following equality
\begin{eqnarray}
S_{A'} = {\rm tr}\log D_{A'} = - \lambda_{+}\log\lambda_{+} - \lambda_{-}\log\lambda_{-}
\end{eqnarray}
where $\lambda_{+}=\cos^2{\theta \over 2}$ and $\lambda_{-}=\sin^2{\theta \over 2}$ are the two eigenvalues of $D_{A'}$. Hence, we get
\begin{eqnarray} \label{entropytheta}
S_{A'}(\theta) = -\cos^2{\theta \over 2} \log\cos^2{\theta \over 2}-\sin^2{\theta \over 2}\log\sin^2{\theta \over 2}
\end{eqnarray}
We know already, because for $\theta = 0$ and $\theta = \pi$, we have $A'$ coinciding respectively with $A_{up}$ and $A_{down}$, that for these two values $S_{A'}(\theta) = 0$, because the states of the ball in these points, namely $D_{(1,0,\phi)}$ and $D_{(1,\pi,\phi)}$ are pure states (see (\ref{theta0}) and (\ref{thetapi})). When we investigate the values of $S_{A'}(\theta)$ for other values of $\theta \in [0, \pi]$, we see that the value increases for increasing values of $\theta$, reaches a maximum for $\theta = {\pi \over 2}$, which is when $A'$ is located in the center of the sphere in Figure \ref{ElasticSphereModel3Dfigure}, and in the middle point of the elastic stretched between $A_{down}$ and $A_{up}$ in Figures \ref{ElasticSphereModelfigure} and  \ref{ElasticSphereModel3Dfigure}, and its maximum value is
\begin{eqnarray}
S_{A'}({\pi \over 2}) &=& -({1 \over \sqrt{2}})^2 \log (({1 \over \sqrt{2}})^2)-({1 \over \sqrt{2}})^2 \log (({1 \over \sqrt{2}})^2) \nonumber \\
&=&- 2 {1 \over 2} \log {1 \over 2} \nonumber \\
&=& \log 2
\end{eqnarray}
which is the maximum possible value able to be reached by the von Neumann entropy in quantum mechanics, and the density operator presenting this state of maximum entropy is given by
\begin{eqnarray}
D_{\rho=0} =
\begin{pmatrix}
{1 \over 2} & 0 \\
0 & {1 \over 2}
\end{pmatrix}
\end{eqnarray}
We now have all the elements to get to the core of this section of the article. Each point $(\rho, \theta, \phi)$,  $\rho \in [0,1]$, $\theta \in [0,\pi]$, $\phi \in [0,2\pi]$, within the sphere shown in Figure \ref{ElasticSphereModel3Dfigure} corresponds to a density operator $D_{(\rho,\theta,\phi)}$, and the points of the sphere surface, i.e., $\rho=1$, correspond to pure states of the quantum entity. The points of the interior of the sphere correspond to density states of the quantum entity. With a measurement corresponds a convex subspace of the interior of the sphere. For a measurement with possible final states after the measurement $A_{up}$ and $A_{down}$, this convex subspace is given by the line segment from $(1,\pi,\phi)$ to $(1,0,\phi)$, where the elastic is stretched. But all lines passing through the center of the sphere in Figure \ref{ElasticSphereModel3Dfigure} harbor such an equivalent measurement. The points on the line all correspond to a density state, except for the end points, which are indeed part of the surface of the sphere, and thus correspond to a pure state. When a measurement commences on an entity that is in a pure state, such as this one in point $A$, a construction is first made of what this state in point $A$ means to the measuring apparatus located on a line between two diametrical points of the sphere surface, such as the points $A_{down}$ and $A_{up}$. That is the meaning of the orthogonal projection which brings the state in $A$ to a state in $A'$. Once the state has become the one corresponding to point $A'$, then the entity is in a mixed state. The transformation from the state corresponding to $A$ to the state corresponding to $A'$ can be read on the matrix representation, namely, it is the non-diagonal terms of the matrix of the density state that disappear to arrive at the state corresponding to $A'$.

In 1995, one of the authors used the properties of the quantum measurement model developed in \citep{aerts1986} to model the decision process of an opinion poll \citep{aertsaerts1995}. The underlying reasoning that led to modeling the decision process of an opinion poll was that, once one has properly understood the internal dynamics of the elastic sphere model that we analyzed in detail above, it is quite natural that in experiments with participants who are human beings, the same kind of fluctuations will be present in many cases. In 2002, we began to study in a more specific way a well-known problem in human cognition in concept theory, namely, the pet-fish problem, containing an effect called the `guppy effect' \citep{gaboraaerts2002,aertsgabora2005a,aertsgabora2005b}. The aspect we used of the elastic measurement model for the guppy effect was the way it allowed context to be explicitly modeled. The measurement context is expressed mathematically by the convex space of density states, and hence the measurement changes an original non-contextualized pure state, such as the state belonging to $A$, to a contextualized density state, which is the state belonging to $A'$ in the case of $A$. 
In a later phase, we also described the interference that could be modeled with more extensive data on the combinations of concepts, by describing this combination with a state that is a quantum superposition of the states that describe the concepts separately \citep{aerts2009a}, see, for example, the interference of
{\it Fruits} and
{\it Vegetables} with data obtained in experiments conducted by James Hampton in the 1980s (Section 3 of \citep{aerts2009b}).

To introduce our study of `human perception', we will use an experiment and its analysis by Jerome Bruner \citep{brunerpostman1949}. Note that Bruner, considered to be one of the most important psychologists of the twentieth century, worked on perception in the 1940s, an era where cognitive science was not yet defined as a subdiscipline in psychology. In the experiment we will describe in detail here, Bruner's intention was to confront participants with stimuli he called `incongruent'. To make clear what this term `incongruent' means, we outline the view on perception in which Bruner framed this experiment. Bruner worked within a view of `perception' in which the hypothesis is advanced that, a perception corresponding to a stimulus is an event in which there is always what he calls a `construction--defense' balance at work. From a pattern of expectations, a structure present in the mind of the one who perceives, `construction' takes place when a stimulus meets an expectation, whereas `defense' takes place when a stimulus counters an expectation. In the experiment we will now describe, participants were confronted with stimuli that confirmed but also contradicted the structure of expectations with which participants' minds took part in the experiment. The purpose of the experiment was to distinguish and analyze different phases of dealing with this contradiction, which Bruner names `incongruity'. In our search for the presence of quantum structures, we are particularly interested in one of the phases, which he calls the `dominance reaction'. Let us now first describe the experiment in \citep{brunerpostman1949} and then further on specify in which way we proceed in identifying a primitive to emerge quantum~structure.

Twenty-eight participants, students at Harvard and Radcliffe, were shown successively by tachistoscopic exposure five different playing cards. One to four of these cards was incongruous---color and suit were reversed. The order of presentation of normal and incongruous cards was randomized. The normal and `trick' cards used were the following. Normal cards (printed in their proper color), five of hearts, ace of hearts, five of spades, and seven of spades. Trick cards (printed with color reversed), three of hearts (black), four of hearts (black), two of spades (red), six of spades (red), ace of diamonds (black), and six of clubs (red). Fourteen orders of presentation were worked out, and two subjects were presented the cards in each of these orders. There were three types of stimulus series, (1)~a single trick card embedded in a series of four normal cards, (2) a single normal card embedded in a series of four trick cards, (3) a mixed series in which trick and normal cards were in the ratio of 3:2 or 2:3. Each card was presented successively until correct recognition occurred, three times each at 10, 30, 50, 70, 100, 150, 200, 250, 300, 400, 450, and 500 ms, and then in steps of 100 ms to 1000 ms. If at 1000 ms, recognition did not occur, the next card was presented. In determining thresholds, correct recognition was defined as two successive correct responses. At each exposure, the subject was asked to report everything he saw or thought he saw. The cards were mounted on medium gray cardboard and were shown in a Dodge--Gerbrands tachistoscope. The pre-exposure field was of the same gray color and consistency as the exposure field, save that it contained no playing card. The light in the tachistoscope was provided by two G E. daylight fluorescent tubes.

For those readers who would like more details about the experiment and its analysis and conclusions, we refer to \citep{brunerpostman1949}. We are particularly interested in the different reactions of the participants, and especially the one that Bruner called the `dominant reaction'. This dominant reaction to the cards shown swiftly consists of providing a response that is entirely within the expectation pattern of the participant, namely, that these are standard cards that have not been tricked. With respect to the trick cards, this reaction consists, essentially, of a `perceptual denial' of the incongruous elements in the stimulus pattern. Faced with a red six of spades, for example, a participant may report with considerable assurance, `the six of spades' or the `six of hearts', depending upon whether he is color or form bound. In the one case, the stimulus form dominates and the color is assimilated to it, in the other, the stimulus color dominates and form is assimilated to it. In both instances, the perceptual resultant conforms with past expectations about the `normal' nature of playing cards.
Let us note that what takes place here, and what is generalizable to any process of `stimulus versus perception', is very similar to what takes place in the experiments with the Necker cube, where the expected image is one of two possible three-dimensional cubes, the one viewed from above or the one viewed from below \citep{conteetal2009,atmanspacherfilk2010}. From this similarity, the hypothesis is justified that if experiments were designed that tested the probability model of this stimulus versus perception interaction, this probability model would prove to be non-classical.

We will not describe the other possible reactions further, as we are concerned with this one, incidentally most common reaction. With longer times of showing the cards, and repeating the experiment with the same participants, a different pattern does develop, with the participants eventually finding out that trick cards are being used---a real repetition of the experiment, to calculate, e.g., the collapse probabilities, should always invite other participants.
If we return to our previous analysis of the quantum measurement model, we can mention some points. The points on the line between $A_{down}$ and $A_{up}$ are a classic `lack of knowledge' space, which makes the state associated with point $A'$ entail a lack of knowledge about which of the two, within the context of the one who measures, $A_{up}$ or $A_{down}$, is the case. That, by the way, is also the reason why the von Neumann entropy of the state associated with $A'$ is different from zero, which stands for that state carrying a lack of knowledge. From the moment $A$ is changed to $A'$, the classical human decision process begins, choosing between $A_{down}$ and $A_{up}$. No choice can be made for a state that is out of context, i.e., outside the points that lie on the line from $A_{down}$ to $A_{up}$. Thus, if we interpret the quantum measurement model in this way, there is not even a way to step `out of the context', the points on the line between $A_{down}$ and $A_{up}$.

In Bruner's work, perception is considered a phenomenon that, in addition to the influence of the stimulus, is also determined by the presence of the expectation pattern of the one who perceives. There is, however, structurally much more to it, partly as a consequence of this basic structure of `perception'. The coming about of the conceptual in itself, as a form of quantization, was experimentally identified, step by step, admittedly. How this developed gradually we bring out in the next section, where we introduce the notion of `categorical perception'.

\section{Categorical Perception and Quantum Structure \label{categoricalquantum}}

In this section, we consider the phenomenon of `categorical perception'. That phenomenon is rooted in the experimental finding that our perception is `warped' so that differences between stimuli we classify in different categories are magnified, while differences between stimuli we classify in the same category are reduced. The first example of categorical perception occurred not so long after Bruner's work on perception, namely, in 1957. A continuum of evenly distributed consonant-vowel syllables with endpoints reliably identified as `b', `d', and `g' was then generated by Liberman and his collaborators. It was observed that there was a point where there was a rapid decrease in the probability of hearing the sound as a `b' compared to hearing it as `d'. At a later point, there was a rapid switch from `d' to `g' \citep{libermanetal1957}. Liberman was also at the origin of what is now called `the motor theory of speech perception'. It is worth elaborating on this briefly.
This theory of speech perception found its origin in work on a speech device developed in the 1950s called the `pattern playback machine', which intended to make it possible to automatically convert texts into spoken form so that, for example, blind people could read with it. For this, it was necessary to analyze speech very thoroughly, and Liberman stumbled upon the phenomenon that would become known as `categorical perception'. The theory rests on the basic hypothesis that people perceive spoken words by recognizing the gestures in the speech channel with which they are uttered, rather than by identifying the sound patterns that produce the speech \citep{libermanetal.1967}. More specifically, the reason why people perceive an abrupt change between `b' and `p' in the way we hear speech sounds is determined by the way people produce these sounds as they speak. Unlike what happens with a synthetic morphing device, a person's natural vocal apparatus is unable to pronounce anything between `b' and `p'. Thus, when someone hears a sound from a synthetic morphing device, that person tries to compare that sound to what he or she would have to do with his or her vocal apparatus to produce this sound. Since a human vocal apparatus can only produce `b' or `p', all continuous synthetic stimuli will be perceived as `b' or `p', whichever is closest.

The theory came under criticism when it was shown that `identification' and `discrimination' of stimuli not connected to speech at all behave in a similar way to those connected to speech, when measured in a similar way \citep{lane1965}. Additionally, it was found that before children could speak, they already exhibited the specific categorical perception effect associated with speech perception that had been identified in adults \citep{eimasetal1971}. It started to become clear that categorical perception was a phenomenon much more general than simply one associated with speech, and the connection was put forward with earlier findings related to the way stimuli are organized. Specifically, Lawrence's experiments and the theory of `acquired distinctiveness' that he formulated from them, which happened in the same year as Bruner's experiment, had revealed what turned out to be a very basic effect of perception. The theory of acquired distinctiveness states that stimuli for which one learns to give distinct responses become more distinct, whereas stimuli for which one learns to give the same response become more similar \citep{lawrence1949}. Both effects are at work in humans across a multitude of perceptions. Stimuli that fall within the same category are perceived as more similar, whereas stimuli that fall into different categories are perceived as more different. What happens with `colors' is a good example of the phenomenon of categorical perception. We see a discrete set of colors, namely, the colors of the rainbow, though in the physical reality there is a continuum of different frequencies offered to us as stimuli. The warping effect of the categorical perception of colors consists of the fact that two stimuli that both fall within the category of `green' are perceived as more similar than two stimuli of which one falls within the category of green and the second within the category of blue, even if from a physics perspective the both pairs of stimuli have the same difference in frequencies. The cooperation of these two effects, a contraction within an existing category, and a dilation between different categories, causes a clustering together into clumps of colors, which in the end gives rise to the seven colors of the rainbow.

Words too are such clumps formed in one way or another from perceptions, with simultaneous contraction and dilation as the warping mechanism of the stimuli. Categorical perception suggests in that sense at least a weak form of the so-called Sapir--Whorf hypothesis, namely, that there is an influence of `how we name categories with language' and `the perception of stimuli belonging to these categories'. In a first phase it was believed that this is not the case for colors, colors were found to be universal and not subject to how they were named in different languages. Not only do most cultures divide the clumps of colors in a similar way, and give them separate names, but also, for the few cultures where this is not the case, the regions of compression and dilation were believed to be the same. We all see blue and green in a similar way, with a fuzzy part in between, even if the naming is not the same \citep{berlinkay1969}. This view was, however, contested by studies that nonetheless identified effects of the words that denote colors.
Comparative research on colors between speakers of Setswana, a Bantu language spoken in South Africa by about 8.2 million people, and speakers of English, found many similarities but also identified differences that are relevant in terms of the Sapir--Whorf hypothesis \citep{daviesetal1998}. Speakers of Berinmo, an indigenous language in Papua New Guinea, have only one word, `nol', for what English speakers call `green' and `blue'. The difference they made compared to speakers of English in color discrimination tasks related to shades between green and blue was examined and determined \citep{davidoff2001}. Later, evidence was found that linguistic categories affect categorical perception mainly in the right visual field. Since the right visual field is controlled by the left hemisphere, this finding was explained by the fact that language skills are also located in the left hemisphere~\citep{regierkay2009}.
Recent sophisticated experiments have demonstrated very thoroughly that language and the names given have a fundamental influence on the categorization that takes place with very primitive visual perceptions. Nine-month-old infants were shown a continua of novel creature-like objects. There was a learning phase, where infants were shown that objects from one end of the perceptual continuum moved left and objects from the other end moved right. For one group of infants, the objects were always named with the same name, whereas for another group of infants, two different names were used to name the objects according to whether they belonged to one end of the continuum or the other. The test consisted of showing all infants new objects from the same continuum, and then seeing if a difference occurred between the two groups. What was found is that infants in the one-name condition formed one overarching category and looked for new test objects in either location. Infants in the two-name condition distinguished two categories and correctly anticipated the likely locations of the test objects, even if they were near the poles or near the center of the continuum \citep{havywaxman2016}. Even if this kind of experiment shows that the phenomenon of categorical perception, i.e., the warping of stimuli, contracting toward sameness for those that will begin to form a category, and dilating toward distinctiveness for those that will belong to different categories, is always at work at all times in learning, there are indeed elements of it that already exist at birth. Our sensory perceptions of stimuli for both color and speech sounds have already been largely warped by evolution in the compression dilation manner inherent in categorical perception. Before we move on to a deeper analysis of what categorical perception may now mean in terms of the presence of quantum structures, let us briefly add to the rich experimental evidence base for the phenomenon by considering the categorical perception of emotions. Emotions are often studied by considering the seeing of facial expressions as stimuli, and it turns out that, like colors and speech sounds, these facial expressions are categorical, and across cultures consist of six basic categories, happiness, disgust, sadness, surprise, anger, and fear \citep{hessetal2009,disaetal2011}.

Is categorical perception merely the mechanism that produces a warping in the collection of stimuli?
In the examples of the situations we considered, we left out an important aspect, namely, that the mechanism of forming a category often also involves stimuli that if analyzed purely physically are totally different. Thus, there is a `similarity' in a `difference' present at a very fundamental level that we must make explicit. For example, all experiments whose intention is to start from basic stimuli in order to identify and quantify the mechanism of warping, actually start from a situation in which major important categorical choices already exist, and are without any doubt innate, namely, those connected with the various senses of human beings.
Indeed, the senses already force a very fundamental categorization, to which practically speaking, all others are subordinate. All the senses, in very fine and deep coordination and agreement, represent to us, among other things, a world in which objects are present and can move, in a space that is three-dimensional, and in which these objects can also interact.
In this world, other people are also present, with bodies like our own and minds like our own. Communicating with those others can be done through language. As we are born with these different senses, a large part of the categorizing of stimuli as `belonging to the same category' consists of stimuli which when analyzed in themselves are of totally different natures, just think of `sounds' and `visual stimuli', which from the everyday experience of the world are `grouped together' in a category. If we want to understand more deeply the rule of `contracting' stimuli that belong to the same category and `dilating' stimuli that belong to different categories, we have to introduce a more general way to consider `similarity'.
It was crucial in terms of survival for our ancestors in the African savanna that they placed `the roar of a lion' as a stimulus in the same category as `the sight of a lion', whereas the former is a sound and the latter a visual stimulus, and thus as stimuli very different in themselves. It was also important that the sound of `the roar of a lion' was considered different from the sound of `the roar of thunder' in a specific way, different, for example, from the difference between `the roar of the lion' and `the sight of the lion'. However, it is not just being born with different senses that makes it necessary to broaden the notion of `similarity'. In \citep{goldstonehendrickson2010}, a good example is given of how far the `similarity' for essentially different stimuli extends when it comes to forming categories. For the category `things to be rescued from a burning house', `children' and `jewelry' are `similar' and members of this category. Introducing, mainly under the influence of the psychologist Murray Sidman, who applied it in his work with mentally retarded children to teach them basic skills, the notion of `equivalence relation' for `similarity', was theoretically a major step forward in understanding what `categorical perception' structurally produces \citep{sidman1994}. An equivalence relation is one which is `reflexive' ($x \sim x$), `symmetrical' ($x \sim y \Rightarrow y \sim x$), and `transitive' ($x \sim y$, $y \sim z \Rightarrow x \sim z$), and if we say that `children' and `jewels' are similar, then it is better to speak rather of `equivalence', specifying at the same time the category with respect to which this equivalence applies, namely, `with respect to things to be saved from a burning house'. This approach which consists of focusing on equivalence relations relative to categories, has already been used to describe aspects of animal social and communicative interactions such as kinship, friendship, coalitions, territorial behavior, and referential calling. In this way, this approach offers a new experiment-based understanding of how animals without language interact with categories in their environment, and more specifically, this has been investigated for sea lions \citep{schustermanetal2000}.

After this passage on the notion of `equivalence relation', and thus a theoretical modeling of what psychologists who study perceptions might call `the collection of stimuli' by means of `the structure of categories', we want to put forward the two next steps of development. The coming into existence of prototype theory as a modelization for concepts is the first step we want to mention \citep{rosch1973}. The basic idea of prototype theory is that for a concept there is a central element, which we call the prototype, relative to which the exemplars of the concept can be placed within a graded structure. Eleanor Rosch put forward the idea that would develop into the primary model for concepts by studying the categorical structure among the Dani for colors and basic shapes.  The Dani are a people living in Papua New Guinea who  possess only two words designating colors, one meaning `bright' and the other `dark'. The Dani also do not possess words in their language for basic shapes, such as circle, square, or triangle. Rosch investigated whether there was a difference in learning between two groups of Dani volunteers, one group was taught colors and basic shapes, starting with stimuli that were prototype colors and prototype basic shapes, and the other group was taught starting with stimuli that were distinct warpings of these prototypes. In a significant way, it appeared that for both colors and basic shapes, learning was more qualitative for the group who was taught starting from the prototype stimuli. This evaluation of `more qualitative learning' took into account the three characteristics of how this can be measured, i.e., by ease of learning sets of form categories when a particular type was the prototype, by ease of learning individual types within sets, and by rank order of judgment of types as best examples of categories, when the prototypes of both colors and shapes were the stimuli in the learning process. Mathematical prototype models based on fuzzy sets were developed for concepts and tested experimentally, and it seemed that the way in which by warping, i.e., contracting when stimuli fall into the same category and dilating when they fall into different categories, concepts emerge and grow from stimuli, had finally found a form in which one can understand what is taking place on a fundamental theoretical level \citep{collieretal1973,rosch1975,roschetal1976,smithmedin1981,medinetal1984,geeraerts2001,johansenkruschke2005}.

However, what was definitely wrong with prototype theory, at least in terms of the idea that a prototype, with a possibly very complex gradual structure of associated exemplars, could really serve as the basic type for any concept, became clear when the combination of two concepts was considered. As we mentioned in the introduction to this article, it was the guppy effect that revealed this problem of the combination of concepts \citep{oshersonsmith1981}. For one of the authors, this guppy effect was also the trigger to try to identify quantum structures in human concepts \citep{gaboraaerts2002,aertsgabora2005a,aertsgabora2005b}. Let us reframe the results of that time, taking into account the knowledge gained in the meantime, the work on the `equivalence relation' that we mentioned above, and especially the basic approach of the present article, namely, `to start from the stimuli', `move through the perceptions', and then `go to the concepts', but also `back to the perceptions'.

One of the authors worked on the elaboration of axiomatic quantum mechanics, in which `states' of entities and `properties' of those entities are the basic elements, and the basic mathematical structure of this axiomatic theory is thus a `state properties  system'~\citep{aerts1982,aerts1992,aerts2002}. This axiomatic quantum mechanics is also an operational theory, meaning that all elements, both states and properties, are operationally defined and introduced.  It is true that the theory is intended to ultimately describe quantum entities, but since all the elements, `states' and `properties', are introduced operationally, `the nature of what the entities are plays no role whatsoever in this axiomatic theory'. This therefore means that the entities considered in this axiomatic theory can also be `concepts', since nowhere is it explicitly assumed that the quantum entities intended to be described by this axiomatic theory would be `objects'. It is only assumed that they are `entities' that are in a specific state under a specific context. This structural similarity was the inspiration for introducing the notion of the `state context property system', or SCoP, as an axiomatic theory for the description of concepts \citep{aertsgabora2005a,aertsgabora2005b}. It was also the first finding at the time regarding the idea that a general quantum formalism, such as what was built in this axiomatic approach \citep{aerts2002}, would be able to provide a satisfactory description of human concepts. It was also the motivation for introducing the notion of the
`state of a concept', and it also seemed to be the right generalization with respect to the old idea that a concept
would be a `collection of exemplars'. Rather conversely, it are `exemplars that are specific states of the concept of which they are exemplars'.

It is important to note that the `introduction of the notion of state' for a concept actually goes back further in the history of how humans made progress in describing reality. Indeed, this `notion of state' was also already used in classical mechanics. It was a deep insight that grew in the period between Galileo and Newton that when describing a material object, the velocity of that object is as important when it comes to grasping its deep reality as it is for the position of that object. We know this intuitively, if we are crossing a busy street with heavy traffic, we need to monitor the velocity at which we cross that street as much as our position if we are to do so safely. Later came the more subtle insight, and it was more specifically Descartes who came up with that proposal that not velocity, but momentum, was the more relevant variable complementary to position. That it is rather the momentum $p$, equal in non-relativistic mechanics to $m$, the mass, multiplied by $v$, the velocity, that is important, we also know intuitively. If we cross that busy street while carrying a heavy weight, e.g., a child on our shoulders, we have to extra monitor our velocity along with our position, because the extra weight makes our movement very different. This is because it is the momentum, to which the weight of our own body plus that of the child we are carrying on our shoulders contribute, that determines how we can move. In the more profound classical mechanics theories, the state of a material object is described by the couple $(x, p)$, where $x$ is the position of the object and $p$ is the momentum of that object. The set of all these couples $(x,p)$ for a material object is called its phase space, and the more abstract theories of classical mechanics, e.g., Hamilton's theory, describe any material object with as its state space its phase space.  It is also, in the context we are considering in this article, interesting to note that the classical mechanics knowledge of this Hamiltonian description of material objects in their phase space was crucial to the emergence of quantum mechanics. It is no coincidence that `the Hamiltonian' is a basic concept in Hilbert space quantum mechanics, it is the operator that represents energy and describes the evolution of the quantum system over time.

In \citep{aertsgabora2005a} we studied how concepts change under different contexts. In doing so, the name `ground state' was given to what is called `the prototype' in prototype theory, and the exemplars in the SCoP formalism are `excited states' of the same concept, which is considered `an entity' that can be in different states. Take the concept {\it Pet}. By it, the guppy effect was in its first phase explained as a context effect. Indeed, different contexts were considered, two of which are given here as examples, the context, `The pet is chewing a bone', and the context, `Did you see the type of pet he has? This explains that he is a weird person'. For the first context, {\it Dog} turned out to be the most typical pet, and for the second context this most typical pet was {\it Spider}. The intention of the SCoP modelization is to allow for dynamic change under the influence of context, following the example of a material object that moves and is also influenced by context, e.g., the gravitational field of the earth, the inertia of mass, the nature of space, and also moving cars on the street, when we consider the example of someone crossing a busy street that was mentioned above. The static picture in which a basic version of prototype theory falls short would, if applied to the example of the person crossing a busy street, consist of one, or several, photographs, of the dynamic scene, and then assuming that these photographs could form the basis for a theory of what is happening. However, the contexts present in the situations described in what is called classical mechanics, such as the gravitational field of the earth, the inertia of material entities, and space, are always very global, and  change little. The cars on the street are contexts that within classical mechanics are already very difficult to grasp in theory, only if special ad hoc models are considered they can be modeled, because even each car of those driving on the street is still a classical context. A context such as, `The pet is chewing a bone', or `Did you see the type of pet he has? This explains that he is a weird person', as considered in \citep{aertsgabora2005a}, is of a much more immediate and intrinsic nature. It is a context built from concepts, and the entity which changes state is itself a concept. This immediacy and intrinsicness is reminiscent of the contextuality that appears in quantum mechanics. It also was an interesting avenue of research to try out whether the quantum formalism would be able to model the contextual dynamics that is clearly and very measurably present in human concepts as entities in a SCoP formalism.

The path taken with the ScoP formalism as inspiration, with a concept as a cloud of different exemplar states in a graded structure around the prototype proved very fruitful.
Let us explain how starting from a Schr\"odinger wave function one comes to such a model. The absolute value squared of this wave function is the probability distribution of localization of the quantum particle described by that Schr\"odinger wave. The typical graded structure, the cloud of exemplars around a prototype, which is traditionally mathematically modeled using fuzzy set theory, can always also be modeled by such a probability distribution derived from a Schr\"odinger wave function, a fuzzy set, and these probability distributions are indeed mathematically isomorphic structures.
However, why is the quantum formalism essentially richer than fuzzy set theory? Well, the probability distribution belonging to a Schr\"odinger wave is `the square of the absolute value' of the Schr\"odinger wave, and this `square of absolute value' makes what is called the phase of the wave disappear, since the absolute value of any phase is equal to 1. This unique state of affairs, namely, that the fuzzy aspect of quantum mechanics is only obtained `after' an absolute value of the Schr\"odinger wave function is squared, determines the intrinsic nature of quantum mechanics. Indeed, the superposition principle, the heart of quantum mechanics, is an operation of summation performed on the Schr\"odinger wave function itself, and `not' on the square of the absolute value of this wave function.

A quantum model was worked out for the guppy effect with a Schr\"odinger wave function \citep{aerts2009b,aertsetal2012}, using data on the combination of concepts collected by James Hampton in the 1980s \citep{hampton1988a,hampton1988b}.
Our first example models  the concepts {\it Fruits} and {\it Vegetables}, and their combination {\it Fruits or Vegetables} with data from \citep{hampton1988b}.
In each case, participants in the experiment provide estimates on a Likert scale of the extent to which a considered exemplar `is' or `is not' a member of one of the three concepts. The data are represented in Table \ref{FruitsVegetablestable}. Hence, concretely, the $\mu(A)_{7}=0.1138$ in the first column of Table \ref{FruitsVegetablestable} is the membership weight the participants attributed to the exemplar {\it Elderberry} for {\it Fruits}, $\mu(B)_{14}=0.0545$ in the second column of Table \ref{FruitsVegetablestable} is the membership weight the participants attributed to the exemplar {\it Mushroom} for {\it Vegetables}, and $\mu(A\ {\rm or}\ B)_{4}=0.0415$ is the membership weight the participants attributed to the exemplar {\it Olive} for {\it Fruits or Vegetables}. Note that the numbers in Table \ref{FruitsVegetablestable} are not the same as the ones Hampton measured. To represent the weights in a Hilbert space requires a renormalization, such that the sum of all the weights equals one. These weights could also be measured directly in an experiment, but one would not use a Likert scale to do so, one would repeatedly pick one of the exemplars as the preferred member. The experiment would be much more cumbersome, but from an experimental standpoint closer to measuring the nature of membership. Hampton's Likert scale values were used to calculate these renormalized membership weights.

\begin{figure}[h!]
\begin{center}
\includegraphics[width=11cm]{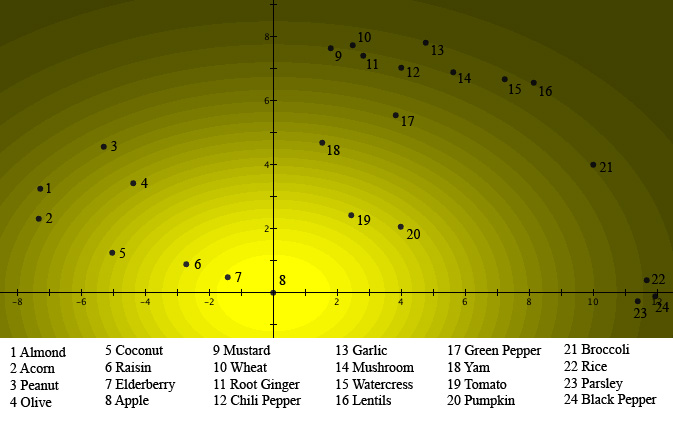}
\end{center}
\caption{The membership weight $\mu(A)_k$ of exemplar $k$ for {\it Fruits} is fitted into a two-dimensional quantum wave function $\psi_A(x,y)$. The numbers are placed at the locations of the different exemplars with the membership weights corresponding to the probability distribution $|\psi_A(x,y)|^2$. This can be seen as a light source shining through a hole centered on the origin, where {\it Apple} is located. The brightness of the light source in a specific region corresponds to the membership weight of this exemplar for {\it Fruits}.}
\label{Fruitsfigure}
\end{figure}

\begin{figure}[h!]
\begin{center}
\includegraphics[width=11cm]{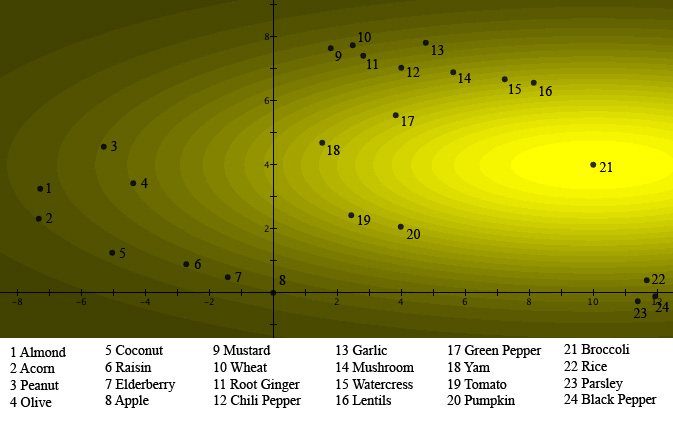}
\end{center}
\caption{The membership weights $\mu(B)_k$ of exemplar $k$ for {\it Vegetables} were fitted into a two-dimensional quantum wave function $\psi_B(x,y)$. The numbers were placed at the locations of the different exemplars with the membership weights corresponding to the probability distribution $|\psi_B(x,y)|^2$. This can be seen as a light source shining through a hole centered on point 21, where {\it Broccoli} is located. The brightness of the light source in a specific region corresponds to the membership weight of this exemplar for {\it Vegetables}.}
\label{Vegetablesfigure}
\end{figure}

The next step consists of representing these data as the squares of the absolute value of three Schr\"odinger wave functions, of which the Schr\"odinger wave function describing the situation for {\it Fruits or Vegetables} is the superposition of the two Schr\"odinger wave functions, one describing the situation of {\it Fruits} and  one describing the situation of {\it Vegetables}. To visualize this quantum model in a 25-dimensional Hilbert space, and to compare it with a situation that has been much studied in quantum physics, namely, the double-slit experiment, the exemplars were given places on a plane, which represents a detection plate. Figure \ref{Fruitsfigure} represents the situation for {\it Fruits}. 
\begin{table}[h!]
\small
\begin{center}
\begin{tabular}{lllllll}
\multicolumn{2}{l}{} & \multicolumn{1}{l}{\boldmath{$\mu(A)_k$}} & \multicolumn{1}{l}{\boldmath{$\mu(B)_k$}} & \multicolumn{1}{l}{\boldmath{$\mu(A\ {\rm or}\ B)_k$}}
& \multicolumn{1}{l}{\boldmath{${1 \over 2}(\mu(A)_k+\mu(B)_k)$}} & \multicolumn{1}{l}{\boldmath{$\phi_k$}} \\
\multicolumn{7}{l}{\it \boldmath{$A$} \textbf{= Fruits,} \boldmath{$B$} \textbf{= Vegetables}} \\
1 & {\it Almond} & 0.0359 & 0.0133 & 0.0269 & 0.0246 & 83.8854$^\circ$ \\
2 & {\it Acorn} & 0.0425 & 0.0108 & 0.0249 & 0.0266 &  $-$94.5520$^\circ$ \\
3 & {\it Peanut} & 0.0372 & 0.0220 & 0.0269 & 0.0296 &  $-$95.3620$^\circ$ \\
4 & {\it Olive} & 0.0586 & 0.0269 & 0.0415 & 0.0428 &  91.8715$^\circ$ \\
5 & {\it Coconut} & 0.0755 & 0.0125 & 0.0604 & 0.0440 & 57.9533$^\circ$ \\
6 & {\it Raisin} & 0.1026 & 0.0170 & 0.0555 & 0.0598 &  95.8648$^\circ$ \\
7 & {\it Elderberry} & 0.1138 & 0.0170 & 0.0480 & 0.0654 & $-$113.2431$^\circ$ \\
8 & {\it Apple} & 0.1184 & 0.0155 & 0.0688 & 0.0670 &  87.6039$^\circ$ \\
9 & {\it Mustard} & 0.0149 & 0.0250 & 0.0146 & 0.0199 &  $-$105.9806$^\circ$ \\
10 & {\it Wheat} & 0.0136 & 0.0255 & 0.0165 & 0.0195 &  99.3810$^\circ$ \\
11 & {\it Root Ginger} & 0.0157 & 0.0323 & 0.0385 & 0.0240 &  50.0889$^\circ$ \\
12 & {\it Chili Pepper} & 0.0167 & 0.0446 & 0.0323 & 0.0306 &   $-$86.4374$^\circ$ \\
13 & {\it Garlic} & 0.0100 & 0.0301 & 0.0293 & 0.0200 & $-$57.6399$^\circ$ \\
14 & {\it Mushroom} & 0.0140 & 0.0545 & 0.0604 & 0.0342 &  18.6744$^\circ$ \\
15 & {\it Watercress} & 0.0112 & 0.0658 & 0.0482 & 0.0385 &  $-$69.0705$^\circ$ \\
16 & {\it Lentils} & 0.0095 & 0.0713 & 0.0338 & 0.0404 & 104.7126$^\circ$ \\
17 & {\it Green Pepper} & 0.0324 & 0.0788 & 0.0506 & 0.0556 &  $-$95.6518$^\circ$ \\
18 & {\it Yam} & 0.0533 & 0.0724 & 0.0541 & 0.0628 &  98.0833$^\circ$ \\
19 & {\it Tomato} & 0.0881 & 0.0679 & 0.0688 & 0.0780 & 100.7557$^\circ$ \\
20 & {\it Pumpkin} & 0.0797 & 0.0713 & 0.0579 & 0.0755 & $-$103.4804$^\circ$  \\
21 & {\it Broccoli} & 0.0143 & 0.1284 & 0.0642 & 0.0713 & $-$99.6048$^\circ$ \\
22 & {\it Rice} & 0.0140 & 0.0412 & 0.0248 & 0.0276 & $-$96.6635$^\circ$ \\
23 & {\it Parsley} & 0.0155 & 0.0266 & 0.0308 & 0.0210 & $-$61.1698$^\circ$ \\
24 & {\it Black Pepper} & 0.0127 & 0.0294 & 0.0222 & 0.0211 &  86.6308$^\circ$ \\
\end{tabular}
\end{center}
\normalsize
\caption{Interference data for concepts $A$, {\it Fruits}, and $B$, {\it Vegetables}. The membership weight of an exemplar $k$ for {\it Fruits} ({\it Vegetables}, respectively) is given by $\mu(A)_k$ ($\mu(B)_k$, respectively). The membership weight of an exemplar $k$ for {\it Fruits or vegetables} is given by $\mu(A\ {\rm or}\ B)_k$. $\phi_k$ is the quantum interference angle.
\label{FruitsVegetablestable}}
\end{table}
The 24 exemplars are all states of {\it Fruits}, and `places' were designated on the detection plate for each of them. Thus, these are the places where quantum particles would be detected when  the `{\it Fruits} slit' is open. Figure \ref{Fruitsfigure} clearly shows the graded structure typical of a prototype model, namely, the square of the absolute value of the Schr\"odinger wave function, describing the situation of {\it Fruits} using Hampton's data. It can be seen that {\it Apple} is at the center of the light source, hence being the prototype for {\it Fruits}. The further away an exemplar is from {\it Apple}, the lower the participants estimated its membership weight. This square of the absolute value of the Schr\"odinger wave function is also equal to the light intensity when a light source shines through the `{\it Fruits} slit'. Underlying this, however, is the Schr\"odinger wave function itself, which possesses a phase that will play a fundamental role when another Schr\"odinger wave function is encountered. This other Schr\"odinger wave function is found in Figure \ref{Vegetablesfigure}, where the data collected by Hampton for the concept of {\it Vegetables} are depicted. This time, the same participants were asked to estimate on a Likert scale how much the same exemplars tested for the concept {\it Fruits} `are' or `are not' members of the concept {\it Vegetables}. Again, this figure clearly shows the graded structure of the prototype model, and indeed, just as was the case for {\it Fruits}, the square of the absolute value of the Schr\"odinger wave function for {\it Vegetables} was used to model Hampton's data. Again, as is the case for {\it Fruits}, we can view this figure as light shining through a `{\it Vegetables} slit', whose particles are then detected on the detection plate. This time, {\it Broccoli} is the center of the light source, hence being the prototype of {\it Vegetables}. The further away an exemplar is from {\it Broccoli}, the lower participants have rated this exemplar on the Likert scale in terms of membership. 
Again, the Schr\"odinger wave function is underlying, with a phase that disappears by squaring the absolute value, but a phase that will play a crucial role if a superposition with another Schr\"odinger wave function is performed.

\begin{figure}[h!]
\begin{center}
\includegraphics[width=11cm]{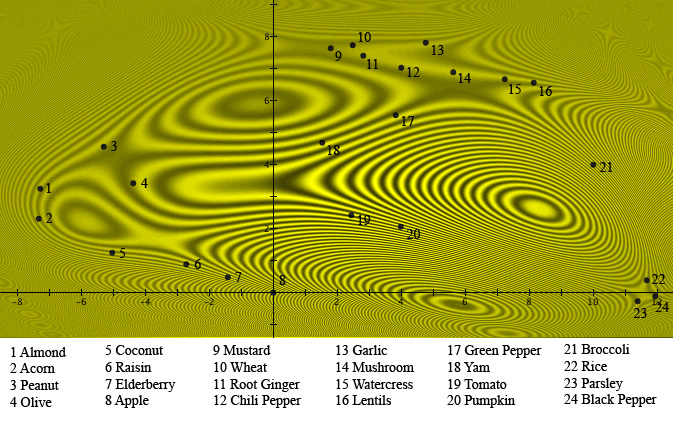}
\end{center}
\caption{The membership weights $\mu(A\ {\rm or}\ B)_k$ of exemplar $k$ for {\it Fruits or Vegetables} were fitted into a two-dimensional quantum wave function $\psi_{A{\rm or}B}(x,y) = {1 \over \sqrt{2}}(\psi_A(x,y)+\psi_B(x,y))$. The numbers were placed at the locations of the different exemplars with the membership weight corresponding to the probability distribution $|\psi_{A{\rm or}B}(x,y)|^2={1 \over 2}(|\psi_A(x,y)|^2+|\psi_B(x,y)|^2)+|\psi_A(x,y)\psi_B(x,y)|\cos\phi(x,y)$, where $\phi(x,y)$ is the quantum phase difference at $(x,y)$. The values of $\phi(x,y)$ are given in Table \ref{FruitsVegetablestable} for the locations of the different exemplars.This can be seen as a light source shining with both the `{\it Fruits} slit' and the `{\it Vegetables} slit' open, and an interference patterns is clearly visible.}
\label{FruitsOrVegetablesfigure}
\end{figure}

Figure \ref{FruitsOrVegetablesfigure} depicts Hampton's data obtained for the concept {\it Fruits or Vegetables}, the combination that is a disjunction of the two concepts {\it Fruits} and {\it Vegetables}. This time we use a Schr\"odinger wave function that is a superposition of the Schr\"odinger wave functions for {\it Fruits} and the Schr\"odinger wave function for {\it Vegetables}. One sees that one no longer gets a simple graded structure but an interference pattern. Indeed, for this composite concept, there is no longer a center of the light source, and it is the situation where both slits, the `{\it Fruits} slit' and the {\it Vegetables slit} are open. Now let us work out the above mathematically. Let is choose the Schr\"odinger wave functions for {\it Fruits}, {\it Vegetables}, and {\it Fruits or Vegetables} as functions of two variables $x$ and $y$, which describe the locations on the detection plate by means of a coordinate system $(x, y)$, and call them $\psi_A(x,y), \psi_B(x,y)$, and $\psi_{A{\rm or}B}(x,y)$, respectively. We chose $\psi_A(x,y)$ and $\psi_B(x,y)$ such that the real part of both wave functions is a Gaussian in two dimensions, such that the top of the first Gaussian is in the origin, and the top of the second Gaussian is located at point $(a,b)$.
Hence,
\begin{eqnarray}
\psi_A(x,y)&=&\sqrt{D_A}e^{-({x^2 \over 4\sigma^2_{Ax}}+{y^2 \over 4\sigma^2_{Ay}})}e^{iS_A(x,y)} \\
\psi_B(x,y)&=&\sqrt{D_B}e^{-({(x-a)^2 \over 4\sigma^2_{Bx}}+{(y-b)^2 \over 4\sigma^2_{By}})}e^{iS_B(x,y)}
\end{eqnarray}
The squares of these Gaussians---also Gaussians---represent the membership weights and are graphically represented in Figures \ref{Fruitsfigure} and \ref{Vegetablesfigure}, and the different exemplars of Table \ref{FruitsVegetablestable} are located in spots such that the Gaussian distributions $|\psi_A(x,y)|^2$ and $|\psi_B(x,y)|^2$ properly model the membership weights $\mu(A)_k$ and $\mu(B)_k$ in Table \ref{FruitsVegetablestable} for each one of the exemplars. This is always possible, taking into account the parameters of the Gaussians, $D_A$, $D_B$, $\sigma_{Ax}$, $\sigma_{Ay}$, $\sigma_{Bx}$, $\sigma_{By}$, $a$, and $b$; and the necessity to fit 24 values, namely, the values of $\mu(A)_k$ and $\mu(B)_k$ for each of the exemplars of Table \ref{FruitsVegetablestable}. We remark that the constraint comes from the exemplars having to be located in exactly the same points of the plane for both Gaussians. Although the solution is an elaborate mathematical calculation, it is also straightforward, so that its details are left to be worked out by the interested reader. 
It is not unique, but different solutions are topologically stretched versions of the one we use in this article, 
 which means that the interference pattern of other solutions is topologically isomorphic to the one we present here. For the Schr\"odinger wave function $\psi_{A{\rm or}B}(x,y)$, the superposition of $\psi_A(x,y)$ and $\psi_B(x,y)$ is taken, hence the renormalized sum of these two~functions.
\begin{eqnarray}
\psi_{A{\rm or}B}(x,y) = {1 \over \sqrt{2}}(\psi_A(x,y)+\psi_B(x,y))
\end{eqnarray}
The square of the absolute value of this function needs to correspond now with the data that Hampton measured for the membership weights for {\it Fruits or Vegetables}.
Let us calculate the square of the absolute value of $\psi_{A{\rm or}B}(x,y)$. One finds
\begin{eqnarray}
|\psi_{A{\rm or}B}(x,y)|^2={1 \over 2}(|\psi_A(x,y)|^2+|\psi_B(x,y)|^2)+|\psi_A(x,y)\psi_B(x,y)|\cos\phi(x,y)
\end{eqnarray}
where $|\psi_A(x,y)$$\psi_B(x,y)|\cos\phi(x,y)$ is the interference term and
\begin{eqnarray}
\phi(x,y)=S_A(x,y)-S_B(x,y)
\end{eqnarray}
is the quantum phase difference at $(x,y)$. A solution was calculated for the function $\phi(x,y)$ as a linear combination of powers of products of $x$ and $y$, electing the lowest powers to attain 25 linear independent functions, with the constraint that the values of $\phi(x,y)$ coincide with the ones in Table \ref{FruitsVegetablestable} for the locations of the different exemplars. Such a solution exists and is unique. The interference pattern in Figure \ref{FruitsOrVegetablesfigure} is a representation of this solution.

\begin{figure}[h!]
\begin{center}
\includegraphics[width=11cm]{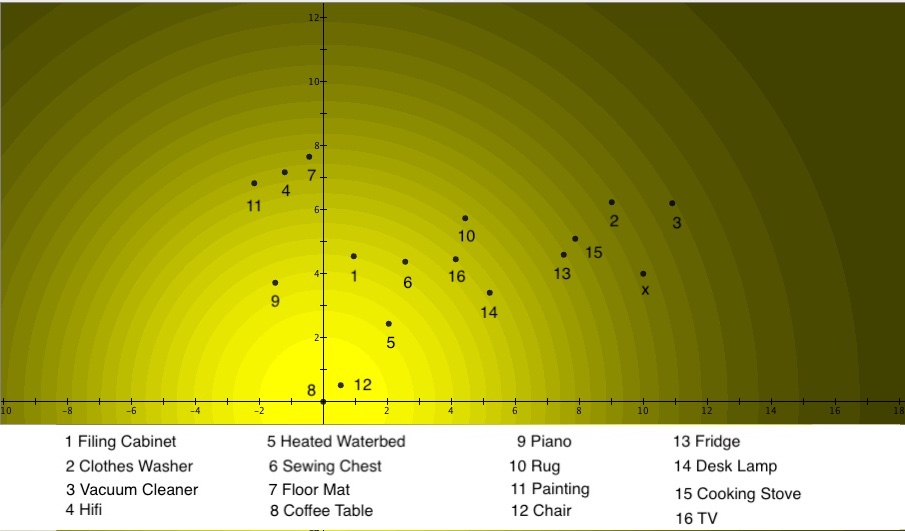}
\end{center}
\caption{The membership weights $\mu(A)_k$ of exemplar $k$ for {\it Furniture} were fitted into a two-dimensional quantum wave function $\psi_A(x,y)$. The numbers were placed at the locations of the different exemplars with the membership weights corresponding to the probability distribution $|\psi_A(x,y)|^2$. This can be seen as a light source shining through a hole centered on the origin, where {\it Coffee Table} is located. The brightness of the light source in a specific region corresponds to the membership weight of this exemplar for {\it Furniture}.}
\label{Furniturefigure}
\end{figure}

\begin{figure}[h!]
\begin{center}
\includegraphics[width=11cm]{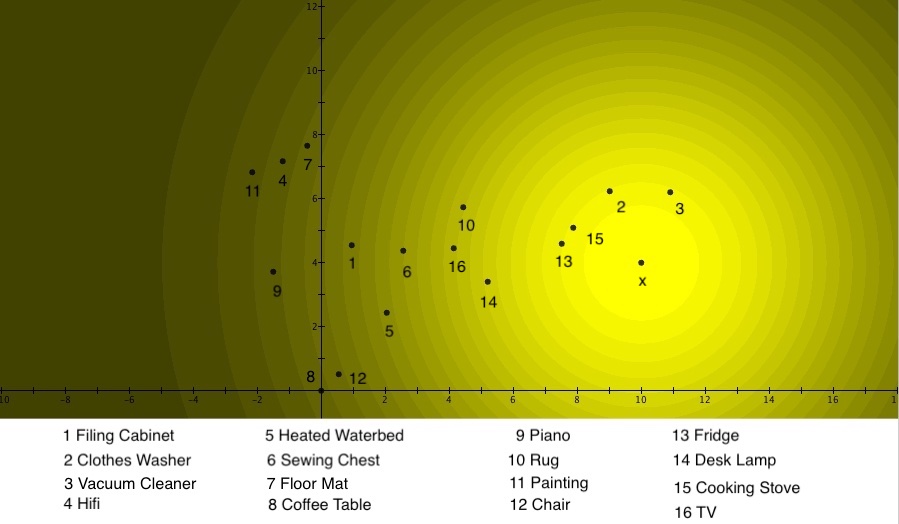}
\end{center}
\caption{The membership weights $\mu(B)_k$ of exemplar $k$ for {\it Household Appliances} were fitted into a two-dimensional quantum wave function $\psi_B(x,y)$. The numbers were placed at the locations of the different exemplars with the membership weights corresponding to the probability distribution $|\psi_B(x,y)|^2$. This can be seen as a light source shining through a hole centered on point $x$. The brightness of the light source in a specific region corresponds to the membership weight of this exemplar for {\it Household Appliances}.}
\label{HouseholdAppliancesfigure}
\end{figure}

\begin{figure}[h!]
\begin{center}
\includegraphics[width=11cm]{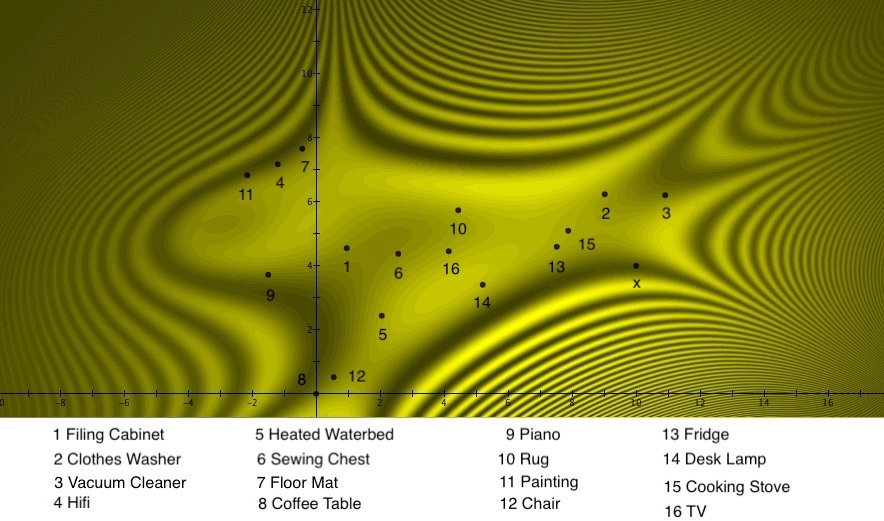}
\end{center}
\caption{The membership weights $\mu(A\ {\rm or}\ B)_k$ of exemplar $k$ for {\it Furniture and Household Appliances} were fitted into a two-dimensional quantum wave function $\psi_{A{\rm or}B}(x,y) = {1 \over \sqrt{2}}(\psi_A(x,y)+\psi_B(x,y))$. The numbers were placed at the locations of the different exemplars with the membership weight corresponding to the probability distribution $|\psi_{A{\rm or}B}(x,y)|^2={1 \over 2}(|\psi_A(x,y)|^2+|\psi_B(x,y)|^2)+|\psi_A(x,y)\psi_B(x,y)|\cos\phi(x,y)$, where $\phi(x,y)$ is the quantum phase difference at $(x,y)$. The values of $\phi(x,y)$ are given in Table \ref{FurnitureHouseholdAppliancestable} for the locations of the different exemplars. This can be seen as a light source shining with both the `{\it Furniture} slit' and the `{\it Household Appliances} slit' open, and an interference patterns is clearly visible.}
\label{FurnitureAndHouseholdAppliancesfigure}
\end{figure}

Our next example models Hampton's data for the concepts {\it Furniture}, {\it Household Appliances}, and {\it Furniture and Household Appliances} in a very similar way as this was the case for {\it Fruits}, {\it Vegetables}, and {\it Fruits or Vegetables} \citep{aertsetal2012}.
While the data for {\it Fruits and Vegetables} involved a composition that is the `disjunction', this time it involves a composition that is a `conjunction'.
This is very similar to the guppy effect. In Figure \ref{Furniturefigure}, one sees the graded pattern for {\it Furniture}, in Figure \ref{HouseholdAppliancesfigure}, the graded pattern for {\it Household Appliances}, and in Figure \ref{FurnitureAndHouseholdAppliancesfigure}, the interference pattern. Instead of describing in detail the images, as was done for the combination of {\it Fruits} and {\it Vegetables}, it is asked `whether one can understand why this kind of interference occurs'. There are two `guppies' among the considered exemplars in Hampton's data, namely, {\it Hifi} and {\it TV}. This can easily be seen in Table \ref{FurnitureHouseholdAppliancestable}. Indeed, the weights for {\it Hifi} are equal to 0.056, 0.076, and 0.090 for {\it Furniture}, {\it Household Appliances}, and {\it Furniture and Household Appliances}, respectively. Thus, the participants in the experiment found that {\it Hifi} is more a member of {\it Furniture and Household Appliances}, than it is a member of {\it Furniture} and than it is a member of {\it Household Appliances}. This is equivalent to {\it Guppy} being more typical of {\it Pet-Fish} than it is typical of {\it Pet} or of {\it Fish}. A similar guppy effect was  found by Hampton for {\it TV}, with values 0.065, 0.092, and 0.099, respectively. 

Can we understand why this interference takes place? Consider how interference takes place, for example, with water waves. If one throws two pebbles into the water, one notices for each pebble how an expanding wave is created in the water. A pebble, where it touches the water, pushes the water molecules into a downward and then upward movement as a reaction. This oscillation of downward and upward motion swirls out in a circle over the water around each of the two pebbles. Where the two expanding ripples of water molecules meet, the typical interference of two water waves occurs. We can easily and intuitively understand what is happening. If the water that was pushed down from one pebble meets the water that is moving up from the other pebble, the two movements cancel each other out, resulting in water that no longer moves. However, if water from the two pebbles meets water from both pebbles that has been pushed down or up, then these movements reinforce each other, and one gets double downward or upward movement in those places. That pattern of annihilating movements on the one hand and amplifying movements on the other is the typical interference pattern.

Can we imagine how a similar phenomenon takes place when concepts are combined? Consider again the example of the `guppy effect'. If {\it Guppy} is confronted with {\it Pet} alone as a concept, and a human mind estimates the size of its membership, a slight upward movement occurs, and likewise if that estimate is made for the membership of {\it Fish} alone. If the same estimate is made for the combination {\it Pet-Fish}, then clearly a much greater upward movement takes place. 
This is a typical `amplifying' case of interference. 
For the {\it Pet-Fish} case, the phenomenon is very easy to sense intuitively. The data that Hampton collected show that similar interference patterns arise that are nevertheless more difficult to sense intuitively. For example, that {\it Hifi} and {\it TV} emerge as the two exemplars with the greatest `guppy' effect is harder to sense intuitively. If one types in {\it Furniture} on Google Images, there is neither a Hifi nor a TV among the images presented, and this is equally true if one types in {\it Household Appliances} on Google Images. 
They are odd ducks in the collection of exemplars Hampton tested in his experiment. Examining Hampton's other data, one can note this as a repeated phenomenon, if an exemplar does not fit well with any of the two concepts, there appears to be a tendency to see that exemplar fit better with a conjunction, but also with a disjunction, of these concepts. Indeed, one might think that the example of the two pebbles thrown into the water and causing an interference pattern only illustrates the conjunction of two concepts, because two pebbles are the conjunction of one pebble with a second pebble. A more detailed study of the combination of concepts, however, where both conjunction and disjunction are considered and compared, and where the negation is also studied, showed that when an exemplar does not fit well at all, neither for one concept nor for the other, the conjunction and disjunction of the two concepts begin to resemble each other \citep{aertssozzoveloz2015a,aertssozzoveloz2015b}. One can also sense this phenomenon intuitively. Consider the example of {\it Olive}, which is an item that is neither a good exemplar of {\it Fruits} nor a good exemplar of {\it Vegetables}. Both the conjunction {\it Fruits and Vegetables} and the disjunction {\it Fruits or Vegetables} welcome {\it Olive} as an exemplar. Indeed, it is appropriate to claim that an olive is somewhat fruit and somewhat vegetable after all, but it is equally appropriate to claim that one is in doubt whether the olive is more of a fruit or a vegetable. After studying many different combinations of concepts, disjunctions, and conjunctions, we arrived at the view that there is an initial preeminent effect which describes the simple combining of the concepts, without logical connectives, i.e., {\it Fruits--Vegetables}, and a second effect, weaker, which represents the logical connectives, the `conjunction' or the `disjunction'. We built a Fock-space model in which these two effects are modeled in two different sectors, and the combination of the two effects is described by a superposition in this Fock-space \citep{aertssozzoveloz2015a}. A stable statistical effect over all tested exemplars of the experiment on which we are relying led us to the hypothesis of the existence of the superposition of two ways of reasoning of the human mind, an emergent one, which is dominant, and the first strongest effect is a consequence of it, and a logical one, which, however, is weaker than the emergent one \citep{aertssozzoveloz2015b}.

We have already mentioned how `words' of human languages, likewise, such as those for colors, can be seen as `clumps' resulting from the warping by the mechanism of categorical perception, and that this warping mechanism is a type of quantization. 
\begin{table}[h!]
\small
\begin{center}
\begin{tabular}{clccccccc}
\multicolumn{2}{l}{} & \multicolumn{1}{l}{\boldmath{$\mu(A)_k$}} & \multicolumn{1}{l}{\boldmath{$\mu(B)_k$}} & \multicolumn{1}{l}{\boldmath{$\mu(A\ {\rm and}\ B)_k$}} & \multicolumn{1}{l}{\boldmath{${1 \over 2}({\mu(A)_k+\mu(B)_k)}$}} &  \multicolumn{1}{l}{\boldmath{$\phi_k$}} && \\
\multicolumn{9}{l}{\it \boldmath{$A$} \textbf{= Furniture,} \boldmath{$B$} \textbf{= Household Appliances}} \\
1&{\it {Filing} Cabinet}&0.079&0.040&0.062&0.059&87.614 &&\\
2&{\it Clothes Washer}&0.026&0.118&0.078&0.072&84.013 & &\\
3&{\it Vacuum Cleaner}&0.017&0,118&0.051&0.068&112.21 & &\\
4&{\it Hifi}&0.056&0.079&0.090&0.067&70.575 & &\\
5&{\it Heated Waterbed}&0.089&0.050&0.082&0.070&79.28 & &\\
6&{\it Sewing Chest}&0.075&0.058&0.061&0.067&94.74 &&\\
7&{\it Floor Mat}&0.052&0.023&0.031&0.037&100.87& &\\
8&{\it Coffee Table}&0,100&0.025&0.050&0.062&104.78&&\\
9&{\it Piano}&0.084&0.020&0.043&0.052&101.67&&\\
10&{\it Rug}&0.056&0.019&0.028&0.037&106.58&&\\
11&{\it Painting}&0.057&0.014&0.021&0.035&120.16&&\\
12&{\it Chair}&0.099&0.030&0.047&0.065&109.41&&\\
13&{\it Fridge}&0.042&0.117&0.085&0.079&85.23&&\\
14&{\it Desk Lamp}&0.066&0.079&0.085&0.072&79.85&&\\
15&{\it Cooking Stove}&0.037&0.118&0.088&0.078&81.57&&\\
16&{\it TV}&0.065&0.092&0.099&0.078&61.89&&\\
\end{tabular}
\end{center}
\caption{Interference data for concepts $A$, {\it Furniture}, and $B$, {\it Household Appliances}. The membership weight of an exemplar $k$ for {\it Furniture} (or {\it Household Appliances}), is given by $\mu(A)_k$ ($\mu(B)_k$, respectively). The membership weight of an exemplar $k$ for {\it Furniture and Household Appliances} is given by $\mu(A\ {\rm and}\ B)_k$. The quantum phase angle $\phi_k$ provokes the quantum interference effect.
\label{FurnitureHouseholdAppliancestable}}
\end{table}
\normalsize
We wish to elaborate on the situation of words as elements of human language from recent work in which one of the authors was involved. From a completely different angle of investigation, it showed that one can consider words structurally as `quanta' of human language. Indeed, an energy scheme can be constructed for texts of stories, and this scheme satisfies the energy distribution of the radiation of a black body, the so-called Planck radiation law~\citep{planck1900}. Moreover, this Planck radiation law for human language had already been noted, purely empirically---without introducing the notion of `energy' admittedly, but bringing in something similar called `ranking'---and called Zipf's law \citep{aertsbeltran2020,aertsbeltran2022}.
In this sense, it is interesting to note that words generally stand for conceptual entities that persist over time. A sentence, or a collection of sentences, uttered during a conversation with a friend, and thus a concatenated string of words, is generally intended to exist only during that brief period when the sentence carries meaning.
Words superpose, just as plane waves superpose mathematically in a Fourier calculus, and therefore cause short-lived phenomena. In this sense, we think that for the study of language as concatenated words, deeply contextual and full of interference, a global quantum formalism in high dimensional Fock spaces is the way to go, and build a performant mathematical formalism.

\section{Conclusions}
We started from a quantum measurement model in which the quantum probabilities arise as a consequence of the existence of fluctuations on the interactions during the measurement itself of the measuring apparatus with the entity to be measured. We have noted how a basic event in human perception, namely, the stimulus against which a perception stands, represents a pure state with a mixture in the context of the expectation pattern of the one who perceives. It is this duality, on the one hand a mixture, within the context of the pattern of expectation, and that is the perception, and on the other hand a pure superposition state, outside the context of the pattern of expectation, and the stimulus existing there, that are captured in its entirety in the measurement model we introduced. From this basis we investigated how the mechanism of categorical perception, which, we believe emerges from this basic duality, mixture and perception, and pure state and stimulus, gives rise to a systematic structuring in a form of perceptual quantization. The most stable concepts can be described in a prototype theory that, however, is unable to capture the fiercely contextual dynamics of even simply compounded concepts in a satisfying way. By introducing an underlying quantum structure, motivated to do so by operational quantum axiomatics, it becomes possible to describe these contextual dynamics of composition using quantum superposition and interference.
Our hypothesis is that the duality of stimulus versus perception, pure superposition state versus mixed state in the expectation space of the one who perceives, is the ground on which this quantum version of a contextual dynamic prototype theory rests. In future research, we wish to bring forward additional evidence for this hypothesis.




\bigskip
\noindent
{\bf Acknowledgments:}
\noindent
We thank the reviewers for interesting comments and suggestions which helped increase the quality of this article. We also thank the members of the Diamond group, Stan Bundervoet, Christian Jendreiko and Marcus Wetzler, parts of the content of this article were presented in their midst on several occasions, and their questions and comments definitely contributed to the realization of our article. We specially want to thank the ArtScience research group of the Center Leo Apostel for Interdisciplinary Studies and its members for the stimulating environment and the regular meetings and discussions offering the fertile ground for our investigation into human perception and its relation to quantum structures. One of us thanks his PhD advisor Francis Heylighen for making it possible to work on this subject of research.This work was supported by QUARTZ (Quantum Information Access and Retrieval Theory), the Marie Sklodowska-Curie Innovation Training Network 7211321 of the European Unions Horizon 2020 research and innovation program and by the Concerted Action Proposal grant ``An Interdisciplinary Study of Creativity: Formal and Empirical Studies of Contextual Effects Across the Arts and Sciences'' of the Vrije Universiteit Brussel.


\end{document}